\documentclass[aps,prd,preprint,
superscriptaddress,preprintnumbers,
showpacs,nofootinbib,nobibnotes]{revtex4}
\usepackage{amsfonts,amssymb,amsmath}
\usepackage{graphicx}
\usepackage{pstricks}
\usepackage{axodraw}

\newcommand{\be}{\begin{equation}}
\newcommand{\bea}{\begin{eqnarray}}
\newcommand{\ee}{\end{equation}}
\newcommand{\eea}{\end{eqnarray}}
\newcommand{\bpi}{\begin{picture}}
\newcommand{\bce}{\begin{center}}
\newcommand{\epi}{\end{picture}}
\newcommand{\ece}{\end{center}}
\newcommand{\D}{\displaystyle}
\def\chic#1{{\scriptscriptstyle #1}}

\def\gb{{\rm I}\hspace{-0.07cm}\Gamma}
\def\g{\widetilde{{\rm I}\hspace{-0.07cm}\Gamma}}

\def\gt{\widetilde{\Gamma}^{\rm L}_{\nu\alpha\beta}}
\def\gnp{{\overline g}^2_{{\chic {\rm NP}}}}
\renewcommand{\theequation}{\arabic{section}.\arabic{equation}}

\newcommand{\Valencia}{Departamento de F\'\i sica Te\'orica
and IFIC, Centro Mixto, Universidad de Valencia--CSIC \\
E-46100, Burjassot, Valencia, Spain}

\begin{document}

%\preprint{FTUV}

\title{Gluon mass generation 
in the PT-BFM scheme}

\author{Arlene~C.~Aguilar}
%\email{aguilar@ift.unesp.br}
\affiliation{Instituto de F\'{\i}sica Te\'orica,
Universidade Estadual Paulista,
Rua Pamplona 145,
01405-900, S\~ao Paulo, SP,
Brazil}

\author{Joannis Papavassiliou}
%\email{Joannis.Papavassiliou@uv.es}
\affiliation{\Valencia}
\date{}

\begin{abstract} 

In this article we study  the general structure and special properties
of the  Schwinger-Dyson equation for the  gluon propagator constructed
with the pinch technique, together  with the question of how to obtain
infrared  finite  solutions,  associated  with the  generation  of  an
effective gluon  mass.  Exploiting the  known all-order correspondence
between  the  pinch technique  and  the  background  field method,  we
demonstrate   that,  contrary   to  the   standard   formulation,  the
non-perturbative gluon self-energy is transverse order-by-order in the
dressed  loop   expansion,  and  separately  for   gluonic  and  ghost
contributions.   We next  present  a comprehensive  review of  several
subtle  issues relevant to  the search  of infrared  finite solutions,
paying  particular attention  to  the  role of  the  seagull graph  in
enforcing transversality, the  necessity of introducing massless poles
in  the  three-gluon vertex,  and  the  incorporation  of the  correct
renormalization group  properties.  In  addition, we present  a method
for  regulating the  seagull-type contributions  based  on dimensional
regularization; its applicability  depends crucially on the asymptotic
behavior of the  solutions in the deep ultraviolet,  and in particular
on the anomalous dimension of the dynamically generated gluon mass.  A
linearized  version  of  the  truncated  Schwinger-Dyson  equation  is
derived, using a vertex that  satisfies the required Ward identity and
contains  massless poles  belonging to  different  Lorentz structures.
The  resulting  integral  equation  is then  solved  numerically,  the
infrared  and ultraviolet  properties  of the  obtained solutions  are
examined in detail, and the allowed range for the effective gluon mass
is determined.   Various open questions and  possible connections with
different approaches in the literature are discussed.

\end{abstract}


\maketitle

\setcounter{section}{0}
\section{Introduction}
\label{Sect:Intro}

The generation  of mass  gaps in  QCD is one  of the  most fundamental
problems in particle physics. In  part the difficulty lies in the fact
that  the  symmetries  governing   the  QCD  Lagrangian  prohibit  the
appearance of  mass terms  for all fundamental  degrees of  freedom at
tree-level and,  provided  that  these symmetries  are  not  violated
through the procedure of regularization, this masslessness persists to
all orders in  perturbation theory.  Thus, mass  generation in QCD
becomes  an   inherently  non-perturbative  problem,   whose  tackling
requires the employment of rather  sophisticated calculational tools
and approximation schemes \cite{Marciano:su}.

Whereas the  generation of quark  masses is intimately  connected with
the breaking of chiral symmetry~\cite{Lane:1974he}, it was argued long
ago that the  non-perturbative QCD dynamics lead to  the generation of
an  effective gluon  mass, while  the  local gauge  invariance of  the
theory                                                          remains
intact~\cite{JMCKursu,Cornwall:1979hz,Parisi:1980jy,Berg:1980gz,
Bhanot:1980fx,Bernard:1981pg,Smit:1974je}.  This gluon ``mass'' is not
a directly measurable quantity, but  must be related to other physical
parameters such  as the  string tension, glueball  masses, or  the QCD
vacuum  energy  \cite{Shifman:1978by},  and  furnishes,  at  least  in
principle, a regulator for all infrared (IR) divergences of QCD.

The concept of a dynamically generated gluon mass, its field theoretic
realization,  and   a  plethora  of  physical   and  technical  issues
associated with  it, have been explored  in great detail  in a classic
paper by Cornwall~\cite{Cornwall:1982zr}.   One of the cornerstones in
his  analysis was  the insistence  on  preserving, at  every level  of
approximation,   crucial    properties   such   as   gauge-invariance,
gauge-independence,  and  invariance  under the  renormalization-group
(RG).    With  this   motivation,  an   effective   gluon  propagator,
$\widehat{\Delta}_{\mu\nu}$,   was  derived  through   the  systematic
rearrangement of Feynman graphs, a  procedure that is now known in the
literature         as          the         ``pinch         technique''
(PT)~\cite{Cornwall:1989gv,Papavassiliou:1989zd,  Binosi:2002ft}.  The
self-energy,   $\widehat{\Pi}_{\mu\nu}$,   of   this   propagator   is
gauge-independent and  captures the leading logarithms  of the theory,
exactly as happens  with the vacuum polarization in  QED.  The central
result   of  ~\cite{Cornwall:1982zr}  was   that,  when   solving  the
Schwinger-Dyson (SD)  equation governing the PT  propagator, and under
special assumptions for the form  of the three-gluon vertex, one finds
solutions that are free of  the Landau singularity, and reach a finite
value in the deep IR.  These solutions may be successfully fitted by a
massive propagator,  with the  crucial characteristic, encoded  in the
corresponding SD equation, that the mass employed is not ``hard'', but
depends non-trivially on the momentum transfer, vanishing sufficiently
fast  in the deep  ultraviolet (UV).   From the  dimensionfull massive
solutions one  may define a dimensionless  quantity, which constitutes
the  generalization   in  a  non-Abelian  context   of  the  universal
(process-independent) QED effective  charge.  The QCD effective charge
so obtained displays  asymptotic freedom in the UV,  whereas in the IR
it ``freezes'' at a finite value.

Various  independent  field  theoretic 
studies~\cite{Bernard:1982my,Lavelle:1991ve,Kondo:2001nq,Bloch:2003yu,Aguilar:2004sw,Dudal:2004rx,Sorella:2006ax}, 
spanning over a quarter of a century,  
also corroborate some type of  gluon mass generation.
In  addition, lattice computations 
\cite{Alexandrou:2000ja,Bonnet:2000kw,Sternbeck:2005tk} reveal the onset of
non-perturbative effects, which in principle can be modelled by means of effectively
massive gluon  propagators.   
It  is
important to  emphasize that the  massive gluon propagator  derived in
~\cite{Cornwall:1982zr}    describes    successfully   nucleon-nucleon
scattering  when  inserted, rather  heuristically,  into the  two-gluon
exchange  model~\cite{Halzen:1992vd}; for  additional phenomenological
applications,    see~\cite{Mihara:2000wf}.     Furthermore,    several
theoretical  studies  based   on  a-priori  very  distinct  approaches
~\cite{Cornwall:1989gv,Mattingly:1993ej,Dokshitzer:1995qm,vonSmekal:1997is,Badalian:1999fq,
Aguilar:2002tc,Brodsky:2002nb,Baldicchi:2002qm,Grunberg:1982fw,Gies:2002af,Shirkov:1997wi,
Gracey:2006dr,Prosperi:2006hx} support the  notion of the ``freezing''
of the QCD running coupling in the deep IR (but do not agree, in
general, on its actual value).

In  recent   years  there  has   been  significant  progress   in  our
understanding of the  PT construction in general~\cite{Binosi:2002ez},
and  the properties of  the resulting  effective Green's  functions in
particular.  The  extension of  the PT to  all orders was  carried out
in~\cite{Binosi:2002ft},  and the known  one-~\cite{Denner:1994nn} and
two-loop~\cite{Papavassiliou:1999az}   connection    with   the   BFM
~\cite{Dewitt:ub,Abbott:1980hw}  was shown to  persist to  all orders.
From the  practical point of  view the established  connection permits
the  direct calculation  with a  set  of concrete  Feynman rules,  and
enables one to prove all-order results, exploiting the powerful formal
machinery of the BFM.  In what  follows we will refer to the framework
emerging from the synergy between PT and BFM as the ``PT-BFM scheme''.

The aim of this article  is threefold: First, we initiate a systematic
treatment  of  the  SD   equations  within  the  PT-BFM  scheme,  with
particular  emphasis on the  manifestly gauge-invariant  truncation it
offers.  Second, we discuss various field-theoretic issues relevant to
the study of gluon mass generation in the context of SD equations in
general.  Third, we analyze in  detail the SD equation obtained as
the first non-trivial approximation in the aforementioned  truncation scheme,
and search for infrared finite solutions.
 
Regarding our first objective, let us point  out that  
one of the most distinct features of the PT-BFM scheme
is the special way in which the transversality of the 
background gluon self-energy
$\widehat{\Pi}_{\mu\nu}$  
is realized. In particular, the study of the 
non-perturbative, SD-type of equation obeyed by $\widehat{\Pi}_{\mu\nu}$
reveals that, 
by virtue of the Abelian-like Ward Identities (WI) 
satisfied by the vertices involved, the transversality 
is preserved 
without the inclusion of ghosts. Put in another way, gluonic
and ghost contributions are {\it separately} transverse.
In addition, transversality is enforced without
mixing the orders in the 
usual ``dressed-loop'' expansion: 
the ``one-loop-dressed'' and ``two-loop-dressed'' sets of 
diagrams are {\it independently} transverse.  
This is to be contrasted to what happens in the usual gauge-fixing scheme 
of the covariant  renormalizable gauges, where the inclusion of the 
ghost is crucial for the transversality already at the level of the one-loop  
perturbative calculation.
This particular transversality property of the BFM self-energy is known at the level of the 
one-loop calculation~\cite{Abbott:1980hw}; however, to the best of our knowledge, 
its all-order generalization  presented in Sec.~\ref{Sect:PT-BFM}  
appears for the first time in the literature. 
The importance of this property in the context of SD equation is that 
it allows for a meaningful first approximation: 
instead of the system of coupled 
equations involving gluon and ghost propagators, one may consider only the 
subset containing gluons, 
without compromising the crucial property of transversality.
More generally, one can envisage a systematic dressed loop expansion,
maintaining transversality manifest at every level of approximation.

Instrumental for some of the developments mentioned above
has been a set of 
non-trivial identities~\cite{Gambino:1999ai}, relating the BFM
$n$-point  functions  to   the  corresponding  conventional  $n$-point
functions  in the covariant  renormalizable gauges, to all  orders in
perturbation   theory.
These  identities, to be referred to as 
Background-Quantum identities (BQIs)~\cite{Binosi:2002ez}, 
are expected to play a fundamental role in addressing one of the most central issues in the context of the 
PT-BFM scheme, namely the actual construction of a new SD series.
Specifically,
as is known already from the two-loop analysis~\cite{Papavassiliou:1999az}, 
the PT-BFM gluon self-energy $\widehat{\Pi}_{\mu\nu}$ is expressed in terms of 
Feynman diagrams containing the conventional  gluon self-energy ${\Pi}_{\mu\nu}$;
this fact is generic, as the all-order diagrammatic representation of $\widehat{\Pi}_{\mu\nu}$ demonstrates
(see Sec.~\ref{Sect:PT-BFM}). Clearly, in order to arrive at a genuine SD equation for 
$\widehat{\Pi}_{\mu\nu}$, one must carry out the substitution 
${\Pi}_{\mu\nu}\to  \widehat{\Pi}_{\mu\nu}$ inside the loops. It is still an open
question whether such a replacement can be implemented self-consistently to all orders;
a preliminary view of how this might work out is presented in Sec.~\ref{Sect:PT-BFM}.

%%%%%%%%%%%%%%%%%%%%%%%%%%%%%%%%%%%%%%%%%%%%%%%%%%%%%%%
Turning to the analysis of the SD equations and the search for infrared finite solutions,
after setting up the appropriate theoretical stage in Sec.~\ref{Sect:GenCon}, in the next two sections 
we eventually 
study a linearized version of the equation governing the 
PT propagator $\widehat{\Delta}_{\mu\nu}$, 
in the spirit of \cite{Cornwall:1982zr}.
Although several of the techniques developed there
are adopted virtually unchanged in the present work, 
there are important theoretical and phenomenological differences, which we summarize below.

(i)  {\it The  role  of the  ghosts:}  As has  become  clear from  the
detailed  study  of  the   correspondence  between  PT  and  BFM,  the
rearrangement  of graphs  (or sets  of graphs)  implemented by  the PT
generates  {\it dynamically}  the characteristic  ghost sector  of the
BFM~\cite{Binosi:2002ft}.    However,  since  the   original  one-loop
derivation of the PT self-energy ~\cite{Cornwall:1982zr} 
the calculations were carried out in
the  context  of the  ghost-free  light  cone  gauge, the  distinction
between  gluonic and  ghost contributions  was not  so obvious.   As a
result, in the heuristic  derivation of the corresponding SD equation,
all contributions  were treated as  gluonic. This is reflected  in the
fact  that   the  coefficient  multiplying   the  characteristic  term
$\ln(q^2+4m^2(q^2))$ appearing in the  solutions (e.g. the standard RG
logarithm  supplemented with the  non-perturbative mass)  is precisely
$b=11  C_A/48\pi^2  $, namely  the  coefficient  of  the one-loop  QCD
$\beta$ function.  Instead, the PT-BFM correspondence reveals that the
purely  gluonic   contributions  is  ${\tilde   b}=10  C_A/48\pi^2  $.
Needless  to  say,  the point  is  not  so  much the  minor  numerical
discrepancy in  the coefficients multiplying the logs,  but rather the
possibility  that  the  ghost  dynamics  may behave  in  a  completely
different  way in  the IR.  Thus,  whilst the  ghosts will  eventually
furnish   the   missing   $C_A/48\pi^2$   asymptotically,   their   IR
contribution  may  deviate from  the  massive  logarithm given  above,
inducing  qualitative   changes  in  the   form  of  the   full  gluon
self-energy.   Reversing the  argument,  in order  to actually  obtain
solutions   of  the  type   $b\ln(q^2+4m^2(q^2))$  from   the  coupled
gluon-ghost system of SD  equations, a very delicate interplay between
gluons and ghosts must take place.

(ii) {\it  Form of the three-gluon  vertex:} It is  well-known that in
order to  obtain dynamically generated  masses one must allow  for the
presence  of massless poles  in the  corresponding expression  for the
three-gluon vertex~\cite{Jackiw:1973tr}.  The effective vertex used in
the  SD equation  of \cite{Cornwall:1982zr}  was the  bare three-gluon
vertex  obtained  from  the  Lagrangian  of  the  (non-renormalizable)
massive           gauge-invariant           Yang-Mills           model
~\cite{KuniGoto,Forshaw:1998tf};  it contains  kinematic  poles, whose
dimensionality is partially compensated  by the explicit appearance of
a hard mass term in  the numerator.  Instead, we use a gauge-technique
inspired  Ansatz   for  the  vertex~\cite{Salam:1963sa},   which  also
contains kinematic  poles, but their dimensionality  is saturated solely
by appropriate  combinations of the momenta involved,  with no explicit
reference to  mass terms,  thus being closer to what  one might
expect to obtain  within QCD.  We hasten to  emphasize that our Ansatz
for the vertex is completely phenomenological, and is not derived from
any dynamical principle, other than the WI that it satisfies, nor does
it exhaust the possible Lorentz  structures.  What we hope to obtain
by  resorting to  such  a simplified  vertex  is a  manageable SD
equation,  that will  allow  us  to study  in  detail the  complicated
interplay of the various components, and get a feel for the dependence
of the solutions on the form of the vertex used.

(iii)   {\it   Seagull    regularization:}   As   was   explained   in
\cite{Cornwall:1982zr},  the integral  equation describing  gluon mass
generation  is supplemented  by a  non-trivial  constraint, expressing
$\widehat{\Delta}^{-1}(0)$  in   terms  of  (quadratically  divergent)
seagull-like contributions; after  its regularization, this constraint
will  restrict  severely  the   number  of  possible  solutions.   The
phenomenological     ``glueball    regularization''     employed    in
\cite{Cornwall:1982zr}  was based on  an elaborate  connection between
the  seagull  contributions, the  massive  Yang-Mills  model, and  the
finite vacuum  expectation value of  a scalar field  creating glueball
states.  Instead,  the regulation we  introduce in this work  is based
solely   on   dimensional    regularization.    In   particular,   the
non-perturbative  seagull contributions  are regulated  by subtracting
from them  the elementary integral $\int d^4k/k^2$,  which vanishes in
dimensional  regularization.  It  turns out  that this  subtraction is
sufficient to regulate  the expression for $\widehat{\Delta}^{-1}(0)$,
provided  that  the  momentum-dependent mass  vanishes  ``sufficiently
fast''  in the  deep UV.  In turn,  this required  asymptotic behavior
restricts  the values  of  the parameters  appearing  in the  integral
equation.

(iv){\it Type of solutions:} 
For relatively moderate values of $\widehat{\Delta}^{-1}(0)$, which 
at the level of the integral equation is treated as an input,
the type of solutions emerging may be fitted with great accuracy by means 
of a {\it monotonically decreasing} 
dynamical mass and a running coupling, exactly as advocated in \cite{Cornwall:1982zr}.
However, as one decreases $\widehat{\Delta}^{-1}(0)$
beyond a critical value,
a new class of qualitatively different 
solutions begins to emerge. These solutions  
are also finite in the entire range of momentum, but 
they display a sharp increase in the deep IR, and 
the corresponding plateau-like range, associated with the ``freezing'', 
becomes increasingly narrower.

%%%%%%%%%%%%%%%%%%%%%%%%%%%%%%%%%%%%%%%%%%%%%%%%%%%%%%%  
The paper is organized as follows:
In Sec.~\ref{Sect:PT-BFM} we review the BT-BFM scheme,
and the structure of the  all-order gluon self-energy.
In Sec.~\ref{Sect:GenCon} we present some general considerations 
pertinent to the search for IR finite solutions, with particular emphasis 
on the role of transversality, the kinematic poles in the vertex employed,
the restoration of the correct UV behavior at the level of the
SD equation, and the regularization of the seagull terms.  
In Sec.~\ref{Sect:linear_toy} we derive the linearized integral equation
and discuss in detail several of its characteristics.
Sec.~\ref{Sect:NumAn} contains the numerical analysis, focusing particularly on the 
appearance of two types of solutions, as mentioned above.
Finally, in Sec.~\ref{Sect:Concl} we discuss connection of this work with other approaches 
in the literature, outline various possible future directions, and summarize our conclusions.   

%%%%%%%%%%%%%%%%%%%%%%%%%%%%%%%%%%%%%%%%%%%%%%%%%%%%%%%%%%%%%%

\setcounter{equation}{0}
\section{The PT-BFM scheme}
\label{Sect:PT-BFM}

In  this  section  we  study  the structure  of  the  effective  gluon
self-energy  obtained  within  the  PT-BFM  framework.  In  the  first
subsection we  present a brief overview  of the PT  and its connection
with the  BFM. The discussion  presented here is  meant to serve  as a
brief reminder; for  a more complete treatment the  reader is referred
to the extensive  literature on the subject. In  the second subsection
we  present   the  all-order  diagrammatic  structure   of  the  gluon
propagator. In the third subsection  we first derive an elementary WI,
valid in  the ghost sector of  the BFM, and then  demonstrate that, to
all  orders, the  contributions  of  gluonic and  ghost  loops to  the
effective gluon self-energy are {\it separately transverse}.  Finally,
in the last subsection we present a preliminary view of how the PT may
eventually lead to a new SD series.

\subsection{The connection between PT and BFM}

The PT
\cite{Cornwall:1982zr,Cornwall:1989gv} is a well-defined algorithm  
that exploits systematically the symmetries
built into physical observables, such as $S$-matrix elements, in order
to construct  new, effective Green's  functions 
endowed  with  very special properties.
Most importantly,
they are independent  of the gauge-fixing 
parameter,  and satisfy  naive  (ghost-free, QED-like) WIs
instead of the usual Slavnov-Taylor  identities.    
The basic observation, which essentially defines the PT, is that there
exists   a   fundamental  cancellation   
between  sets   of  diagrams  with  different  kinematic
properties,  such  as self-energies,  vertices,  and  boxes.   
This
cancellation  is driven   by  the   underlying
BRST symmetry \cite{Becchi:1976nq},
and is triggered 
when a very particular subset of the longitudinal  momenta  circulating
inside vertex and box diagrams generate out of them 
(by ``pinching'' out internal lines) propagator-like terms. 
The latter are reassigned to 
conventional self-energy graphs, in order  to give rise to the 
aforementioned effective  Green's  functions.

%%%%%%%%%%%%%%%%%%%%%%%%%%%%%%%%%%%%%%%%%

The  longitudinal momenta  responsible  for these diagrammatic
rearrangements stem either {\bf (a)} from the bare gluon propagators 
contained inside the various Feynman diagrams, 
\be
\Delta_{\mu\nu}^{[0]}(k) = 
-\frac{\D i}{\D k^2}
\left[\ g_{\mu\nu} - (1-\xi) \frac{\D k_\mu
k_\nu}{\D k^2}\right] \,,
\label{GluProp}
\ee
and/or {\bf (b)} from the
``pinching  part'' $\Gamma^{{\rm  P}}_{\alpha\mu\nu}(q,p_1,p_2)$ 
appearing
in  the  characteristic  decomposition  of the  bare
three-gluon   vertex $\Gamma^{eab}_{\alpha\mu\nu}=
gf^{eab}\Gamma_{\alpha\mu\nu}$ into 
\cite{Cornwall:1982zr}    
\bea
\Gamma_{\alpha\mu\nu}(q,p_1,p_2)&=&  \Gamma^{{\rm
F}}_{\alpha\mu\nu}(q,p_1,p_2)+ \Gamma^{{\rm
P}}_{\alpha\mu\nu}(q,p_1,p_2),        \nonumber\\ 
\Gamma^{{\rm
F}}_{\alpha\mu\nu}(q,p_1,p_2)&=&(p_1-p_2)_\alpha  g_{\mu\nu}+2q_\nu
g_{\alpha\mu}-2q_\mu g_{\alpha\nu},\nonumber    \\    
\Gamma^{{\rm
P}}_{\alpha\mu\nu}(q,p_1,p_2)&=& p_{2\nu}g_{\alpha\mu}-
p_{1\mu}g_{\alpha\nu}.
\label{PTDEC}
\eea 

The case of the gluon self-energy is of particular interest.
Defining the transverse projector
\be
{\rm P}_{\mu\nu}(q)= \ g_{\mu\nu} - \frac{\D q_\mu
q_\nu}{\D q^2}\,,
\label{projector}
\ee
we have for the full gluon propagator in the Feynman gauge
\begin{equation}
\Delta_{\mu\nu}(q)= {-\D i}\left[{\rm P}_{\mu\nu}(q)\Delta(q^2) + 
\frac{q_{\mu}q_{\nu}}{q^4}\right]\,.
\label{prop_cov}
\end{equation}
The scalar function $\Delta(q^2)$ is related to the 
all-order gluon self-energy $\Pi_{\mu\nu}(q)$,
\be
\Pi_{\mu\nu}(q)={\rm P}_{\mu\nu}(q)\Pi(q^2)\,,
\ee
through
\be
\Delta(q^2) = \frac{1}{q^2 + i\Pi(q^2)}\,.
\label{fprog}
\ee
Notice that the way $\Pi_{\mu\nu}(q)$ has been defined in (\ref{fprog}) 
(e.g. with the imaginary  factor $i$ in front), it is   
given simply by the corresponding Feynman diagrams in Minkowski space.
The inverse of the full gluon propagator
has the form
\begin{equation}
\Delta^{-1}_{\mu\nu}(q)=
 i{\rm P}_{\mu\nu}(q) \Delta^{-1}(q^2) + iq_{\mu}q_{\nu}\,,
\label{inv_prog}
\end{equation}
or, equivalently,
\begin{equation}
\Delta^{-1}_{\mu\nu}(q)= ig_{\mu\nu}q^2 - \Pi_{\mu\nu}(q)\,.
\label{inv_prop_pi}
\end{equation}

\begin{figure}[t]
\includegraphics[scale=1.05]{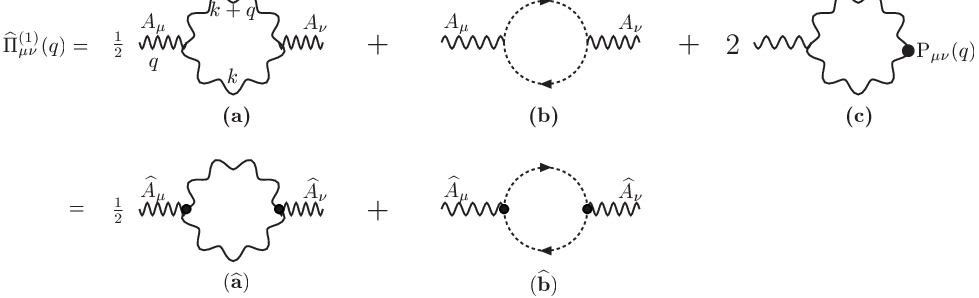}
\caption{The PT self-energy at one-loop.}
\label{oneloopPT}
\end{figure}

The PT construction of the effective one-loop self-energy $\widehat\Pi_{\mu\nu}^{[1]}(q)$ 
can be most easily constructed directly in the Feynman gauge.
It amounts to adding to the conventional 
one-loop $\Pi_{\mu\nu}^{[1]}(q)$ 
( Fig.~\ref{oneloopPT}, ${\bf (a)}$ and ${\bf (b)}$) the pinch contributions   
coming from vertex graphs, shown schematically in ${\bf (c)}$.
%%%%%%%%%%%%%%%%%%%%%%%%%%%%%%%%%%%%%%%%%%%%%%%%%%%%%%%%%%%%%%%%%%%
%%%%%%%%%%%%%%%%%%%%%%%%%FIG. 1%%%%%%%%%%%%%%%%%%%%%%%%%%%%%%%%%%%% 
%%%%%%%%%%%%%%%%%%%%%%%%%%%%%%%%%%%%%%%%%%%%%%%%%%%%%%%%%%%%%%%%%%%%%
%%%%%%%%%%%%%%%%%%%%%%%%%%%%%%%%%%%%%%%%%%%%%%%%%%%%%%%%%%%%%%%%%%%%%%%
Then, the final result is
\be 
\widehat\Delta^{-1}(q^2)= q^2\left[1+ b g^2\ln\left(\frac{q^2}{\mu^2}\right)\right]\,,
\label{rightRG}
\ee
where  $b = 11 C_A/48\pi^2$  is the first coefficient of the QCD $\beta$-function. 

Evidently,
due to the Abelian WIs satisfied by the PT effective Green's functions, the 
 new propagator-like quantity $\widehat\Delta^{-1}(q^2)$ absorbs all  
the RG-logs, exactly as happens in QED with the photon self-energy.
Equivalently, since $Z_{g}$ and ${\widehat Z}_{A}$, the renormalization constants 
of the gauge-coupling and the effective self-energy, respectively, 
satisfy the QED relation ${Z}_{g} = {\widehat Z}^{-1/2}_{A}$,  
the product 
${\widehat d}(q^2) = g^2 \widehat\Delta(q^2)$ forms a RG-invariant 
($\mu$-independent) quantity~\cite{Watson:1996fg,Binosi:2002vk};
for large momenta $q^2$,
\be
{\widehat d}(q^2) = \frac{\overline{g}^2(q^2)}{q^2}\,,
\label{ddef1}
\ee
where $\overline{g}^2(q^2)$ is the RG-invariant effective charge of QCD,
\be
\overline{g}^2(q^2) = \frac{g^2}{1+  b g^2\ln\left(q^2/\mu^2\right)}
= \frac{1}{b\ln\left(q^2/\Lambda^2\right)}\,.
\label{effch}
\ee
Of central importance for what follows is the connection between the PT and the BFM.
The latter is  a  special gauge-fixing 
procedure, implemented  at the  level of  the generating functional.  In 
particular, it preserves  the symmetry of  the action under ordinary  gauge
transformations  with respect to  the background (classical) gauge field
$\widehat{A}^a_{\mu}$,  while the quantum gauge fields $A^a_{\mu}$ appearing in
the  loops transform  homogeneously under  the gauge  group,  {\it i.e.},
as ordinary matter  fields which happened  
to be assigned to  the adjoint representation 
\cite{Weinberg:kr}.   As a  result  of the  background gauge symmetry, the 
$n$-point  functions  $\langle 0  | T \left[ \widehat{A}^{a_1}_{\mu_1}(x_1)
\widehat{A}^{a_2}_{\mu_2}(x_2)\dots  
\widehat{A}^{a_n}_{\mu_n}(x_n) \right] |0 \rangle$
are gauge-invariant, in the sense that they 
satisfy naive, QED-like WIs.
Notice, however, that they are {\it not} gauge-independent, 
because they 
depend {\it explicitly} on the quantum gauge-fixing parameter 
$\xi_Q$ used  to define the tree-level  propagators of  the 
quantum gluons and the three- and four-gluon vertices 
involving one and two background gluons, respectively 
\cite{Abbott:1980hw}.
The connection between  PT and BFM may be stated as follows: 
The (gauge-independent) PT effective $n$-point functions ($n=2,3,4$) 
coincide with the 
(gauge-dependent) BFM $n$-point  functions ($n=2,3,4$ background gluons $\widehat{A}^a_{\mu}$ entering) 
provided that the latter 
are computed at $\xi_Q =1$ 
(e.g. setting $\xi_Q =1$ in the Feynman rules of the Appendix).
This connection was first established at one-loop level~\cite{Denner:1994nn}, 
and was recently
shown to persist to all orders in perturbation theory~\cite{Binosi:2002ft}.

%%%%%%%%%%%%%%%%%%%%%%%%%%%%%%%%%%%%%%%%%%%%%%%%%%%%%%%%%%%
\subsection{The SD equation of the effective gluon self-energy}

The structure of the effective gluon self-energy, as it
emerges from the all-order PT-BFM correspondence,
 can be written in 
a closed non-perturbative form, which coincides with the 
SD equation for $\widehat{\Delta}$, derived formally 
from the BFM path integral using functional techniques~\cite{Sohn:1985em}.

In what follows we assume dimensional regularization, and employ the short-hand notation
$ [dk] =  d^d k/(2\pi)^d$ \,, 
where $d=4-\epsilon$ is the dimension of space-time.
We refer to diagrams containing one explicit integration over virtual
momenta as ``one-loop dressed'' and those with two integrations as ``two-loop dressed''.

We will classify the corresponding diagrams into four categories:
one-loop dressed gluonic contribution (group a), one-loop dressed ghost contribution (group b),
two-loop dressed gluonic contribution (group c), and 
two-loop dressed ghost contribution (group d). 

%%%%%%%%%%%%%%%%%%%%%%%% FIGURE 2 %%%%%%%%%%%%%%%%%%%%%%%%%%%%%%%%%%%%
%      One-loop dressed gluonic contribution - Group A
%%%%%%%%%%%%%%%%%%%%%%%%%%%%%%%%%%%%%%%%%%%%%%%%%%%%%%%%%%%%%%%%%%%%%
\begin{figure}[ht]
\includegraphics[scale=0.8]{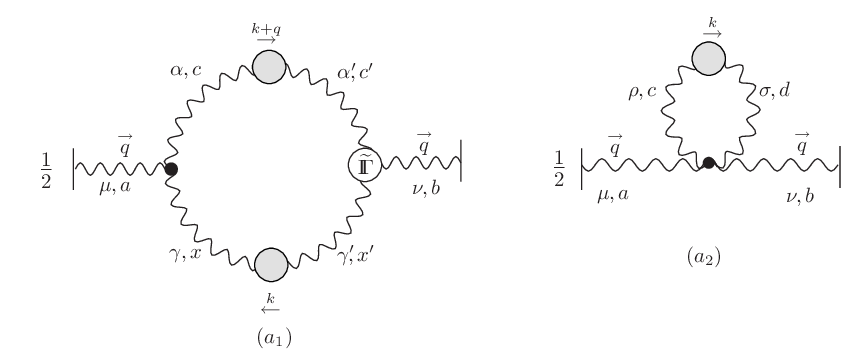}
\caption{The  gluonic  contribution  at
one-loop dressed  expansion. Wavy lines with  grey blobs represent
full-quantum gluon propagators. All external wavy lines (ending with a
vertical line) are background gluons.  The black dots are
the  tree-level  vertices  in  the  BFM, and  the
white-blob denote the full three-gluon vertex with one  background gluon.}
\label{group_a}
\end{figure}
%%%%%%%%%%%%%%%%%%%%%%%%%%%%%%%%%%%%%%%%%%%%%%%%%%%%%%%%%%%%%%%%%%%%%%%
%
%\vspace{1.0cm}
%
The closed expressions corresponding to the two diagrams of Fig.~\ref{group_a} are given by
\begin{eqnarray}
\widehat{\Pi}^{ab}_{\mu\nu}(q)
\big|_{{\bf a_1}} &=& 
\frac{1}{2} \, \int\!\! [dk]\,
\widetilde{\Gamma}_{\mu\alpha\beta}^{aex}
\Delta^{\alpha\rho}_{ee'}(k)
{\g}_{\nu\rho\sigma}^{be'x'}
\Delta^{\beta\sigma}_{xx'}(k+q)\,,
\nonumber\\
\widehat{\Pi}^{ab}_{\mu\nu}(q)
\big|_{{\bf a_2}} &=&
\frac{1}{2} \,\int\!\! [dk]\,
\widetilde{\Gamma}_{\mu\nu\alpha\beta}^{abex}
\Delta^{\alpha\beta}_{ex} (k)\,.
\label{groupa}
\end{eqnarray}

%%%%%%%%%%%%%%%%%%%%%%%% FIGURE 3 %%%%%%%%%%%%%%%%%%%%%%%%%%%%%%%%%%%%%
%           One-loop dressed ghost contribution - Group B
%%%%%%%%%%%%%%%%%%%%%%%%%%%%%%%%%%%%%%%%%%%%%%%%%%%%%%%%%%%%%%%%%%%%
\begin{figure}[ht]
\includegraphics[scale=0.8]{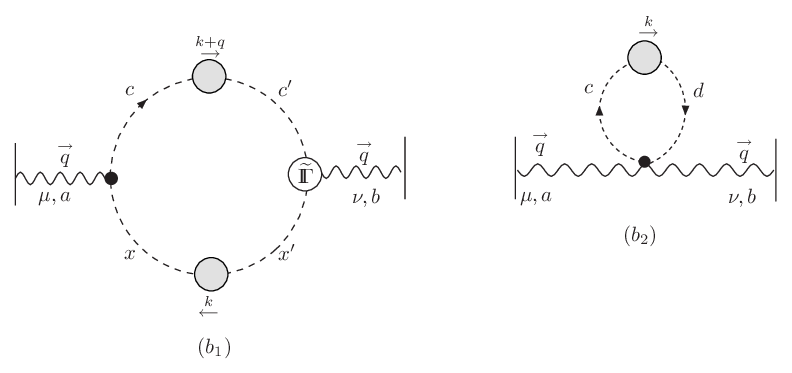}
\caption{The ghost sector at one-loop dressed expansion. Dashed lines with 
grey blobs denote full-ghost propagators, while the white blob 
represents the full background gluon-ghost vertex}
\label{group_b}
\end{figure}
%%%%%%%%%%%%%%%%%%%%%%%%%%%%%%%%%%%%%%%%%%%%%%%%%%%%%%%%%%%%%%%%%%%%%%%

For the Fig.~\ref{group_b} we have
\begin{eqnarray}
\widehat{\Pi}^{ab}_{\mu\nu} (q)
\big|_{{\bf b_1}} &=& 
- \, \int\!\! [dk]\,
\widetilde{\Gamma}_{\mu}^{aex}
D_{ee'}(k) 
{\g}_{\nu}^{be'x'}
D_{xx'}(k+q)\,,
\nonumber\\
\widehat{\Pi}^{ab}_{\mu\nu}(q)
\big|_{{\bf b_2}} &=&
- \int\!\! [dk]\,
\, \widetilde{\Gamma}_{\mu\nu}^{abex}
D_{ex}(k)\,.
\label{groupb}
\end{eqnarray}

%%%%%%%%%%%%%%%%%%%%%%%% FIGURE 4 %%%%%%%%%%%%%%%%%%%%%%%%%%%%%%%%%%%%%
%           Two-loop dressed gluonic contribution - Group C
%%%%%%%%%%%%%%%%%%%%%%%%%%%%%%%%%%%%%%%%%%%%%%%%%%%%%%%%%%%%%%%%%%%%%
\begin{figure}[ht]
\includegraphics[scale=0.8]{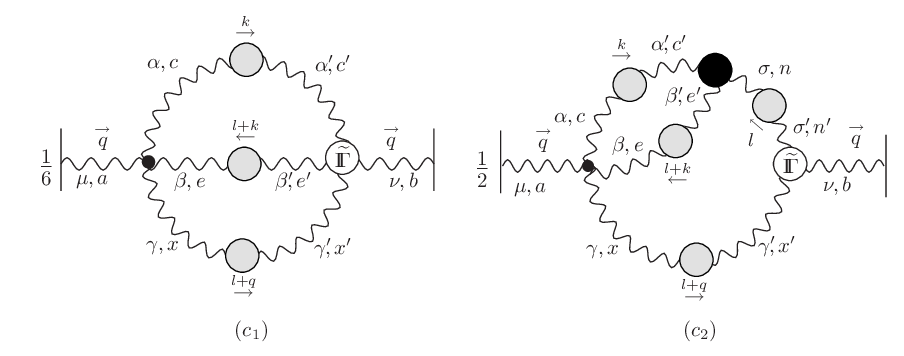}
\caption{Purely gluonic graphs relevant for the two-loop dressed expansion. The black blob represents the
  full conventional three-gluon vertex, while the white blobs 
denote three or four-gluon vertices with one external background leg.}
\label{group_c}
\end{figure}
%%%%%%%%%%%%%%%%%%%%%%%%%%%%%%%%%%%%%%%%%%%%%%%%%%%%%%%%%%%%%%%%%%%%%%%
The two-loop dressed gluonic contribution, Fig.~\ref{group_c}, reads
\begin{eqnarray}
\widehat{\Pi}^{ab}_{\mu\nu}(q)
\big|_{{\bf c_1}} &=& 
\frac{1}{6} \, \int\!\! \int\!\! [dk][d\ell]\,
\widetilde{\Gamma}_{\mu\alpha\beta\gamma}^{acex}
\Delta^{\alpha\alpha'}_{cc'}(k)
\Delta^{\beta\beta'}_{ee'}(k+\ell)
\Delta^{\gamma\gamma'}_{xx'} (\ell+q)
{\g}_{\nu\gamma'\beta'\alpha'}^{bx'e'c'}\,,
\nonumber\\
\widehat{\Pi}^{ab}_{\mu\nu}(q)
\big|_{{\bf c_2}} &=&
\frac{1}{2} \,
\, \int\!\! \int\!\! [dk][d\ell]\,
\widetilde{\Gamma}_{\mu\alpha\beta\gamma}^{acex}
\Delta^{\alpha\alpha'}_{cc'} (k)
\Delta^{\beta\beta'}_{ee'} (k+\ell)
{\gb}_{\sigma\beta'\alpha'}^{ne'c'}
\Delta^{\sigma\sigma'}_{nn'} (\ell)
{\g}_{\nu\gamma'\sigma}^{bx'n'}
\Delta^{\gamma\gamma'}_{xx'}(\ell+q)\,.\nonumber\\
&&{}
\label{groupc}
\end{eqnarray}

%%%%%%%%%%%%%%%%%%%%%%%% FIGURE 5 %%%%%%%%%%%%%%%%%%%%%%%%%%%%%%%%%%%%%
%    Two-loop dressed -  gluon and ghost contribution   Group D
%%%%%%%%%%%%%%%%%%%%%%%%%%%%%%%%%%%%%%%%%%%%%%%%%%%%%%%%%%%%%%%%%%%%%%%
\begin{figure}[ht]
\includegraphics[scale=0.8]{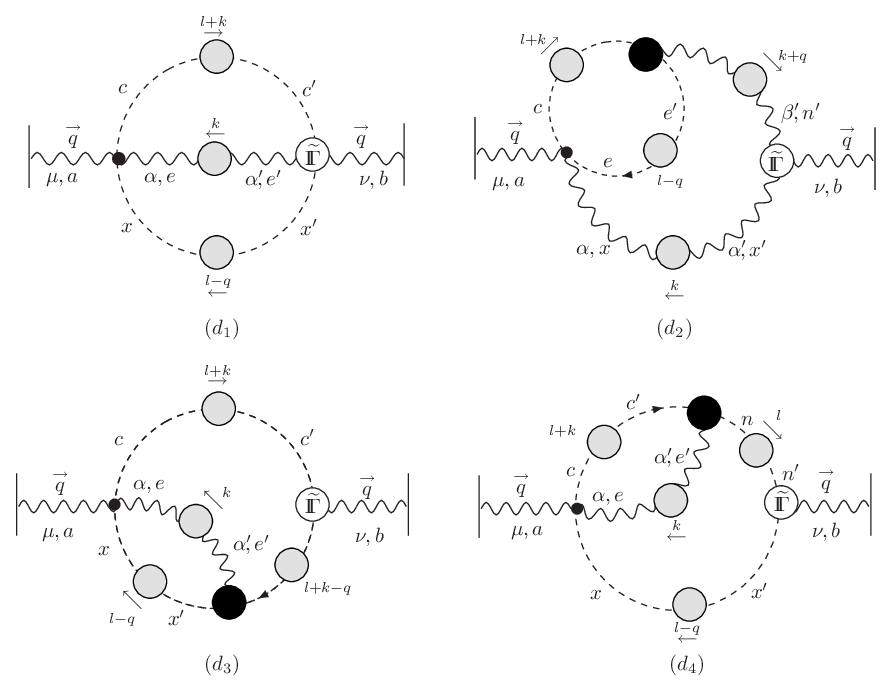}
\caption{The ghost sector contribution to the two-loop dressed expansion. The
black blobs are the conventional full gluon-ghost vertices, while the white ones represent
ghost vertices with an (external) background gluon and two ghosts.}
\label{group_d}
\end{figure}
%%%%%%%%%%%%%%%%%%%%%%%%%%%%%%%%%%%%%%%%%%%%%%%%%%%%%%%%%%%%%%%%%%%%%%%
The last group represents the two-loop dressed ghost contribution, Fig.~\ref{group_d}, and is  written as
\begin{eqnarray}
\widehat{\Pi}^{ab}_{\mu\nu}(q)
\big|_{{\bf d_1}} &=& 
- \int\!\! \int\!\! [dk][d\ell]\,
\widetilde{\Gamma}_{\mu\alpha}^{acex}
D_{cc'}(k+\ell) 
\Delta^{\alpha\alpha'}_{ee'}(k)
D_{xx'} (\ell-q)
{\g}_{\nu\alpha'}^{bx'e'c'}\,,
\nonumber\\
\widehat{\Pi}^{ab}_{\mu\nu}(q)
\big|_{{\bf d_2}} &=&
- \,
\, \int\!\! \int\!\! [dk][d\ell]\,
\widetilde{\Gamma}_{\mu\alpha}^{acex}
D_{cc'}(k+\ell) D_{ee'}(\ell -q)
{\gb}_{\beta}^{e'nc'}
\Delta^{\beta\beta'}_{nn'} (k+q)
{\g}_{\nu\alpha'\beta'}^{bx'n'}
\Delta^{\alpha\alpha'}_{xx'} (k)\,,
\nonumber\\
\widehat{\Pi}^{ab}_{\mu\nu}(q)
\big|_{{\bf d_3}} &=&
- \,
\, \int\!\! \int\!\! [dk][d\ell]\,
\widetilde{\Gamma}_{\mu\alpha}^{acex}
D_{xx'}(\ell -q)\Delta^{\alpha\alpha'}_{ee'} (k)
{\gb}_{\alpha'}^{x'e'n}
D_{cc'}(k+\ell)
{\g}_{\nu}^{bn'c'}
D_{nn'}(k+\ell-q) \,,
\nonumber\\
\widehat{\Pi}^{ab}_{\mu\nu}(q)
\big|_{{\bf d_4}} &=&
- \,
\, \int\!\! \int\!\! [dk][d\ell]\,
\widetilde{\Gamma}_{\mu\alpha}^{acex}
D_{cc'}(k+\ell) \Delta^{\alpha\alpha'}_{ee'} (k)
{\gb}_{\alpha'}^{ne'c'}
D_{nn'}(\ell)
{\g}_{\nu}^{bx'n'}
D_{xx'}(\ell -q)\,.
\label{groupd}
\end{eqnarray}

Notice that, (i) as explained in the Introduction, the propagators appearing inside 
the loops are quantum ones, and (ii) there are two general types of vertices,
those where all incoming fields are quantum, and those where one of the incoming
fields is background.

\subsection{Special transversality properties}

It is well-known that in the conventional formulation, the diagram containing the ghost-loop
(graph ${\bf (b)}$ in Fig.~\ref{oneloopPT}) is instrumental for the transversality 
of $\Pi_{\mu\nu}(q)$. On the other hand, in the 
PT-BFM scheme, due to the special Feynman rules (see Appendix), the contributions of 
graphs  ${\bf (\widehat a)}$ and ${\bf (\widehat b)}$ are {\it individually} transverse. 
Specifically, keeping only the logarithmic terms, one has~\cite{Abbott:1980hw} 
\be 
\widehat\Pi_{\mu\nu}^{{\bf (\widehat a)}}(q) 
= \frac{10\, C_A }{48\pi^2} g^2 \ln\left(\frac{q^2}{\mu^2}\right) 
{\rm P}_{\mu\nu}(q) \,,\,\,\,\,
\widehat\Pi_{\mu\nu}^{{\bf (\widehat b)}}(q) 
= \frac{C_A }{48\pi^2} g^2 \ln\left(\frac{q^2}{\mu^2}\right) 
{\rm P}_{\mu\nu}(q) \,.
\label{split}
\ee

In this subsection we will show that, by virtue of the all-order WI satisfied by the 
full vertices appearing in the diagrams defining  $\widehat\Pi_{\mu\nu}(q)$, Figs.(\ref{group_a}-\ref{group_d}),
 the above property is valid non-perturbatively, and 
that gluonic and ghost contributions are separately transverse. In addition, the 
one-loop and two-loop dressed diagrams do not mix. This is to be contrasted to what happens 
to the conventional case, where the orders of the loop expansion also mix.

There are four fully dressed vertices with one incoming background gluon appearing in the 
diagrammatic definition of $\widehat\Pi_{\mu\nu}(q)$, in Figs.(\ref{group_a}-\ref{group_d}):
${\g}_{\mu\alpha\beta}^{abc}$, ${\g}_{\mu}^{acb}$, ${\g}_{\mu\nu\alpha\beta}^{abcd}$,
${\g}_{\mu\nu}^{cdba}$. As is known from formal considerations (see \cite{Gambino:1999ai}, 
and references therein), the WI obtained when contracting such vertices with the momentum carried
by the background gluon retain to all-orders the same form as at tree-level.
The tree-level WI for ${\g}_{\mu\alpha\beta}^{abc}$ and ${\g}_{\mu}^{acb}$ are simply 
\be
q^{\mu}\widetilde{\Gamma}_{\mu\alpha\beta}^{abc}(q,p_1,p_2) = 
g f^{abc} (p_1^2  - p_2^2 ) g_{\alpha\beta}\,,\,\,\,\
q^{\mu}\widetilde{\Gamma}_{\mu}^{acb}(q,p_1,p_2) = g f^{abc} (p_1^2  - p_2^2 )\,. 
\label{WI2B}
\ee
In addition, the fact 
that the four-gluon vertex with one incoming   background gluon, 
and the conventional 
one (four quantum gluons) coincide at tree-level, as shown in Eq.(\ref{4gluon}),
furnishes the corresponding tree-level
WI for ${\g}_{\mu\nu\alpha\beta}^{abcd}$ (see, for instance~\cite{Papavassiliou:1992ia}).
Therefore, the only ingredient missing is the 
tree-level identity satisfied by ${\g}_{\mu\nu}^{cdba}$; we now proceed to its derivation.

%%%%%%%%%%%%%%%%%%%%%%%%%%%%%%%%%%%%%%%%%%%%%%%%%
Contracting the bare vertex 
$\widetilde\Gamma_{\mu\nu}^{cdba} = -ig^2  f^{acx}f^{xdb} 
g_{\mu\nu}$, shown in Fig.~\ref{bfm_v1},
with the momentum $q_{1}$ carried by the background gluon, 
we have that 
\be
q_1^{\mu}
\widetilde\Gamma_{\mu\nu}^{cdba}(q_{1},q_{2},q_{3},q_{4})
= -ig^2 f^{acx}f^{xdb} q_{1\nu}
= ig^2 \left(f^{abx}f^{cdx} + f^{adx}f^{bcx}\right)q_{1\nu}\,, 
\label{WI4}
\ee
where we have used the Jaccobi identity
\be
f^{abx}f^{cdx} + f^{acx}f^{dbx} + f^{adx}f^{bcx} = 0\,. 
\label{Jac}
\ee
Next we use that 
\be 
0 = - \left(f^{abx}f^{cdx} + f^{acx}f^{dbx} + f^{adx}f^{bcx}\right)
(q_1+q_4)_{\nu}\,,
\ee
and add it by parts to (\ref{WI4}), obtaining 
\bea
q_1^{\mu}
\widetilde\Gamma_{\mu\nu}^{cdba}
&=& ig^2 \left[f^{cdx} f^{axb} q_{4\nu}
+ f^{cbx} f^{adx} q_{4\nu}
+ f^{cax} f^{xdb}(q_1+q_4)_{\nu}\right]\nonumber\\
&=& -ig \left[
f^{cdx} \Gamma_{\nu}^{axb}(q_4,q_2+q_1,q_3) + 
f^{cbx} \Gamma_{\nu}^{adx}(q_4,q_2,q_3 +q_1)+
f^{cax} \Gamma_{\nu}^{xdb} (q_4+q_1,q_2,q_3)\right]\,. 
\nonumber\\
&&{}
\label{WI4b}
\eea

Armed with the above tree-level results, we proceed to state the four fundamental all-order WIs.
First, the WI of the three-field vertices, where on the RHS we have
differences of inverse propagators, are given by
\bea
q_1^{\mu}{\g}_{\mu\alpha\beta}^{abc}(q_1,q_2,q_3) &=&
gf^{abc}
\left[\Delta^{-1}_{\alpha\beta}(q_2)
- \Delta^{-1}_{\alpha\beta}(q_3)\right]\,,
\nonumber\\
q_1^{\mu}{\g}_{\mu}^{acb}(q_2,q_1,q_3) &=&  gf^{abc}
\left[D^{-1}(q_2)- D^{-1}(q_3)\right]\,.
\label{3gl} 
\eea
Then, the WI of the four-field vertices, where on the RHS we have 
sums of three trilinear vertices, with appropriately shifted arguments, are
\bea
q_1^{\mu}{\g}_{\mu\nu\alpha\beta}^{abcd}(q_1,q_2,q_3,q_4) = &&
ig f^{abx} {\gb}_{\alpha\beta\nu}^{cdx}(q_3,q_4,q_1+q_2)
\nonumber\\
 &+&
ig f^{acx} {\gb}_{\beta\nu\alpha}^{dbx}(q_4,q_2,q_1+q_3)
\nonumber\\
 &+&
ig f^{adx} {\gb}_{\nu\alpha\beta}^{bcx}(q_2,q_3,q_1+q_4)\,,
\label{4gl}
\eea
and
\bea
q_{1}^{\mu}
{\g}_{\mu\nu}^{cdba}(q_1,q_2,q_3,q_4) = &&
-ig f^{cdx} \gb_{\nu}^{axb}(q_4,q_2+q_1,q_3)
\nonumber\\
&&
-ig f^{cbx} \gb_{\nu}^{adx}(q_4,q_2,q_3 +q_1)
\nonumber\\
&&
-ig f^{cax} \gb_{\nu}^{xdb} (q_4+q_1,q_2,q_3)\,.
\label{4gh}
\eea
Notice that Eq.(\ref{4gh}) is the 
all-order generalization of  
(\ref{WI4b}).

With the above WI we can prove that the four groups presented 
before are independently transverse. We start with group (a):
we contract graph (${\bf a_1}$) using the first all-order WI of (\ref{3gl}), 
whereas for (${\bf a_2}$) we simply compute the divergence of the tree-level vertex (\ref{bfm4gluon}),
arriving at
\bea
q^{\nu} \widehat{\Pi}^{ab}_{\mu\nu}(q)
\big|_{{\bf a_1}} &=&
 C_A \, g^2 \delta^{ab} \,q_{\mu}  \, \int\!\! [dk]\,
\Delta^{\rho}_{\rho}(k)\,,
\nonumber\\
q^{\nu}\widehat{\Pi}^{ab}_{\mu\nu}(q)
\big|_{{\bf a_2}} &=&
- C_A \, g^2  \delta^{ab} \,q_{\mu}  \, \int\!\! [dk]\,
\Delta^{\rho}_{\rho}(k)\,,
\label{Transgroupa}
\eea
Thus,
\be
q^{\nu}\left( 
\widehat{\Pi}^{ab}_{\mu\nu}(q)\big|_{{\bf a_1}}+
\widehat{\Pi}^{ab}_{\mu\nu}(q)\big|_{{\bf a_2}}\right) = 0\,.
\ee

Similarly, the one-loop-dressed ghost contributions of group (b) give upon
contraction
\bea
q^{\nu} \widehat{\Pi}^{ab}_{\mu\nu}(q)
\big|_{{\bf b_1}} &=&
2 \,C_A \, g^2 \delta^{ab} \,q_{\nu}  \, \int\!\! [dk]\,D(k)\,,
\nonumber\\
q^{\nu} \widehat{\Pi}^{ab}_{\mu\nu}(q)
\big|_{{\bf b_2}} &=& -2 \,C_A \, g^2 \delta^{ab} 
\,q_{\nu}  \, \int\!\! [dk]\,D(k)\,.
\label{Transgroupb}
\eea
and so
\be
q^{\nu}\left( 
\widehat{\Pi}^{ab}_{\mu\nu}(q)\big|_{{\bf b_1}}+
\widehat{\Pi}^{ab}_{\mu\nu}(q)\big|_{{\bf b_2}}\right) = 0\,.
\ee
The two-loop dressed demonstration is slightly more involved, but
essentially straightforward.
We begin with the two-loop gluonic contributions (group c).
The action of $q^{\nu}$ on the all-order four-gluon vertex 
${\g}_{\nu\gamma'\beta'\alpha'}^{bx'e'c'}(-q,\ell+q, -k-\ell,k)$ 
appearing on the RHS of the first equation in (\ref{groupc})
may be 
obtained from  (\ref{4gl}), through the following 
definition of momenta, 
$q_1 = -q$, 
$q_2 = \ell+q$,
$q_3 = -k-\ell$,
$q_4 = k$,  
and corresponding relabellings of Lorentz and color indices. 
Then,
\bea
q^{\nu} \widehat{\Pi}^{ab}_{\mu\nu}(q)
\big|_{{\bf c_1}} &=&
\frac{1}{6} \,ig  \, \int\!\! \int\!\! [dk][d\ell]\,
\widetilde{\Gamma}_{\mu\alpha\beta\gamma}^{acex}
\Delta^{\alpha\alpha'}(k)
\Delta^{\beta\beta'}(k+\ell)
\Delta^{\gamma\gamma'} (\ell+q)[
\nonumber\\
&&
f^{bxi} {\gb}_{\beta'\alpha'\gamma'}^{eci}
(-k-\ell,k,\ell)
+
f^{bei} {\gb}_{\alpha'\gamma'\beta'}^{cxi}
(k,\ell + q ,-k - \ell - q)
+
\nonumber\\
&&
f^{bci} {\gb}_{\gamma'\beta'\alpha'}^{xei}
(\ell + q,-k - \ell, k-q)]\,.
\label{c1}
\nonumber\\
\eea
It is not difficult to verify that, after judicious 
shifting and relabelling of the integration momenta and of the  
``dummy'' Lorentz and color indices, the three terms on the RHS of  
(\ref{c1}) are in fact equal. Thus, 
\be
q^{\nu} \widehat{\Pi}^{ab}_{\mu\nu}(q)
\big|_{{\bf c_1}} = 
\frac{1}{2} \,ig f^{bxi}
\, \int\!\! \int\!\! [dk][d\ell]\,
\widetilde{\Gamma}_{\mu\alpha\beta\gamma}^{acex}
\Delta^{\alpha\alpha'}(k)
\Delta^{\beta\beta'}(k+\ell)
\Delta^{\gamma\gamma'} (\ell+q) 
{\gb}_{\beta'\alpha'\gamma'}^{eci}
(-k-\ell,k,\ell)\,.
\ee
The other graph gives
\bea
q^{\nu} \widehat{\Pi}^{ab}_{\mu\nu}(q)
\big|_{{\bf c_2}} &=&
\frac{1}{2} \,ig \, f^{bxi}
\, \int\!\! \int\!\! [dk][d\ell]\,
\widetilde{\Gamma}_{\mu\alpha\beta\gamma}^{acex}
\Delta^{\alpha\alpha'} (k)
\Delta^{\beta\beta'} (k+\ell)
{\gb}_{\gamma'\beta'\alpha'}^{iec}(\ell,-k-\ell,k)
\Delta^{\gamma'\gamma}(\ell+q)
\nonumber\\
 &-&
\frac{1}{2} \,ig \, f^{bxi}
\, \int\!\! \int\!\! [dk][d\ell]\,
\widetilde{\Gamma}_{\mu\alpha\beta\gamma}^{acex}
\Delta^{\alpha\alpha'} (k)
\Delta^{\beta\beta'} (k+\ell)
{\gb}_{\gamma'\beta'\alpha'}^{iec}(\ell,-k-\ell,k)
\Delta^{\gamma'\gamma} (\ell)\,.
\nonumber\\
&&{}
\label{c2}
\eea
Evidently the second term on the RHS of (\ref{c2}) vanishes identically,
since the integral is independent of $q$, and therefore 
the free Lorentz index $\mu$ cannot be saturated.  

We then observe  that, due to the full Bose symmetry of the conventional 
three-gluon vertex, 
${\gb}_{\beta'\alpha'\gamma'}^{eci}(-k-\ell,k,\ell)
={\gb}_{\gamma'\beta'\alpha'}^{iec}(\ell,-k-\ell,k)$,
and therefore, finally,
\be
q^{\nu}\left( 
\widehat{\Pi}^{ab}_{\mu\nu}(q)\big|_{{\bf c_1}}+
\widehat{\Pi}^{ab}_{\mu\nu}(q)\big|_{{\bf c_2}}\right) = 0\,.
\ee

Finally, we turn to the two-loop-dressed ghost-graphs (group d).
For the calculation of the divergence of graph (${\bf d_1}$)  
we use Eq.(\ref{4gh}); this WI generates three distinct terms, i.e. 
\bea
q^{\nu} \widehat{\Pi}^{ab}_{\mu\nu}(q)
\big|_{{\bf d_1}} &=&
- \,ig  \, \int\!\! \int\!\! [dk][d\ell]\,
\widetilde{\Gamma}_{\mu\alpha}^{acex}
D (k+\ell) 
\Delta^{\alpha\beta} (k)\,
D (\ell-q)[
\nonumber\\
&&
f^{ebi} {\gb}_{\beta}^{xic}
(q-\ell,-k - q,k+\ell)
+
f^{bci} {\gb}_{\beta}^{xei}
(q - \ell, -k, k+\ell - q)
\nonumber\\
&+&
f^{bxi} {\gb}_{\beta}^{iec}
(-\ell,-k,k+\ell) ]\,. 
\label{d1} \,
\nonumber\\
\eea
Each one of these three terms can   
be easily shown to cancel exactly against the 
individual divergences of the remaining three graphs. 
To see this in detail, use the first WI of 
(\ref{3gl}) for graph $({\bf d_2})$, and the second of (\ref{3gl}) 
for  graphs $({\bf d_3})$ and $({\bf d_4})$. In all three cases 
one of the two inverse propagators generated by the WI
will give rise to an expression similar to the second term on the 
RHS of (\ref{c2}), i.e. a $q$-independent integral with a free Lorentz index,
which cannot be saturated. These three terms are directly set to zero.
The terms stemming from the other inverse propagator read:
\bea
q^{\nu} \widehat{\Pi}^{ab}_{\mu\nu}(q)
\big|_{{\bf d_2}} &=& 
\,ig f^{bei} \, \int\!\! \int\!\! [dk][d\ell]\,
\widetilde{\Gamma}_{\mu\alpha}^{acex}\,
D (k+\ell) \Delta^{\alpha\beta} (k)\,
D (\ell-q) {\gb}_{\beta}^{cxi}
(k+\ell,q-\ell,-k - q)\,,
\nonumber\\
q^{\nu} \widehat{\Pi}^{ab}_{\mu\nu}(q)
\big|_{{\bf d_3}} &=&
\,ig   f^{bci}\, \int\!\! \int\!\! [dk][d\ell]\,
\widetilde{\Gamma}_{\mu\alpha}^{acex}\,
D (k+\ell) \Delta^{\alpha\beta} (k)\,
D (\ell-q){\gb}_{\beta}^{ixe}
(\ell + k -q,q - \ell, -k)\,,
\nonumber\\
q^{\nu} \widehat{\Pi}^{ab}_{\mu\nu}(q)
\big|_{{\bf d_4}} &=& 
\,ig f^{bxi} \, \int\!\! \int\!\! [dk][d\ell]\,
\widetilde{\Gamma}_{\mu\alpha}^{acex}\,
D (k+\ell) \Delta^{\alpha\beta} (k)\,
D (\ell-q) {\gb}_{\beta}^{cie}
(k+\ell,-\ell,-k)\,. 
\nonumber\\
&&{}
\label{d234}
\eea
and, as announced, 
can be directly identified with the corresponding terms on the 
RHS of (\ref{c1}). Therefore, 
\be
q^{\nu}\left( 
\widehat{\Pi}^{ab}_{\mu\nu}(q)\big|_{{\bf d_1}}+
\widehat{\Pi}^{ab}_{\mu\nu}(q)\big|_{{\bf d_2}}+ 
\widehat{\Pi}^{ab}_{\mu\nu}(q)\big|_{{\bf d_3}}+
\widehat{\Pi}^{ab}_{\mu\nu}(q)\big|_{{\bf d_4}}\right)= 0\,.
\ee
This concludes the non-perturbative proof of the special transversality property of 
$\widehat{\Pi}^{ab}_{\mu\nu}(q)$: gluon and ghost loops are separately transverse, 
and the loops of different order do not mix

\subsection{Towards a new SD series}

As explained  for the first time in  \cite{Cornwall:1982zr}, the upshot
the PT is to  eventually trade the conventional SD series
for another,  written in terms of the  new, gauge-independent building
blocks.  Then  one could truncate this  new series, by  keeping only a
few  terms  in  a   ``dressed-loop''  expansion,  and  maintain  exact
gauge-invariance,    while   at    the    same   time    accommodating
non-perturbative   effects.    
As mentioned in the Introduction, 
one of the most central issues in this context is how to convert the 
SD-series defining $\widehat{\Delta}_{\mu\nu}$ into a
{\it dynamical} equation, namely one that contains 
$\widehat{\Delta}_{\mu\nu}$ on {\it both} sides. 
For this to become possible, 
one must carry out inside the loops of the diagrams shown in the previous subsection 
the substitution ${\Delta}_{\mu\nu}\to  \widehat{\Delta}_{\mu\nu}$.
In  this   subsection   we  focus only on one particular aspect of this 
problem, namely the role that the BQIs might play in implementing the aforementioned 
substitution.

The connection~\cite{Binosi:2002ez} 
between   the   PT   and   the
Batalin-Vilkovisky quantization formalism \cite{Batalin:pb} has been
instrumental for various of the recent developments in the PT.
Specifically, using  the formulation of the BFM within this latter formalism,  
one can derive non-trivial identities (BQI's), relating the BFM
$n$-point  functions  to   the  corresponding  conventional  $n$-point
functions  in the covariant  renormalizable gauges,  to all  orders in
perturbation   theory~\cite{Gambino:1999ai}.
The relation  between these two types of $n$-point  functions is written in a closed form
by means of a set of auxiliary Green's functions involving
anti-fields   and   background   sources,   introduced   in   the   BFM
formulation. These latter Green's functions are in turn related 
 by  means of a SD  type of equation ~\cite{Binosi:2002ez} 
to the conventional  ghost  Green's  functions  
appearing  in  the STI
satisfied by the conventional all-order three-gluon vertex 
\cite{Bar-Gadda:1979cz,Ball:1980ax,PasTar}.

%%%%%%%%%%%%%%%%%%%%%%%%%%%%%%%%%%%%%%%%%%%%%%%%%%%%%%%%%%%%%%%%%%%
We will restrict our discussion to the case of the propagators $\widehat{\Delta}_{\mu\nu}$ and ${\Delta}_{\mu\nu}$.
The relevant quantity appearing in the corresponding BQI 
is the following two-point function, to be
denoted by $\Lambda_{\alpha \beta}(q)$, defined as  
(we suppress color indices)
\be
\Lambda_{\alpha \beta}(q) = 
\int\![dk]
H^{(0)}_{\alpha\mu}
D(k)\Delta^{\mu\nu}(k+q)\, H_{\beta\nu}^{*}(k+q,-k,-q),
\label{gpert2}
\ee
where the elementary vertex $H_{\alpha\beta}^{(0)}$ is 
%%%%%%%%%%%%%%%%%%%%%%%%%%%%%%%%%%%%%%%%%%%%%%%%%%%%%
%                   figure
%%%%%%%%%%%%%%%%%%%%%%%%%%%%%%%%%%%%%%%%%%%%%%%%%%%%%%
\be
\bpi(0,50)(55,-20)
\PhotonArc(40,0)(25,40,180){-1.5}{6.5}
\DashCArc(40,0)(25,180,320){1}
\DashArrowLine(40.5,-25)(39.5,-25){1}
\Vertex(15,0){1.8}

\Text(5,-1)[r]{$H_{\alpha\beta}^{(0)}\,=$}
\Text(20,0)[l]{$\scriptstyle{\alpha}$}
\Text(62,18)[l]{$\scriptstyle{\beta}$}
\Text(125,-1)[r]{$=\,-igg_{\alpha\beta}$}

\epi
\nonumber
\ee
%%%%%%%%%%%%%%%%%%%%%%%%%%%%%%%%%%%%%%%%%%%%%%%%%%%%%%%%
and  $H_{\alpha\beta}$ is given by
%%%%%%%%%%%%%%%%%%%%%%%%%%%%%%%%%%%%%%%%%%%%%%%%%%%%%%%%
\be
%\bpi(0,160)(75,-120)
\bpi(0,70)(75,-25)

\PhotonArc(140,0)(25,0,180){-1.5}{8.5}
\DashCArc(140,0)(25,180,360){1}
\DashArrowLine(195,-30)(170,-5){1}
\Photon(170,5)(195,30){1.5}{4}
\GCirc(165,0){12}{0.8}
\GCirc(140,23.5){12}{0.8}
\GCirc(140,-25){10}{0.8}
\DashArrowLine(119.2,-13.7)(118.2,-12.3){1}
\DashArrowLine(157.2,-17.8)(156.2,-18.8){1}
\Vertex(115,0){1.8}

\Text(-20,-1)[r]{$H_{\alpha\beta}(p,r,q)$}
\Text(5,-1)[r]{$=$}
\Text(40,-1)[r]{$H_{\alpha\beta}^{(0)}$}
\Text(70,-1)[r]{$+$}
\Text(166,0)[c]{$\scriptstyle{{\cal K}_{\nu\beta}}$}
\Text(140,-25)[c]{$\scriptstyle{D}$}
\Text(140,23.5)[c]{$\scriptstyle{\Delta_{\mu\nu}}$}
\Text(120,0)[l]{$\scriptstyle{\alpha}$}
\Text(198,30)[l]{$\scriptstyle{\beta}$}
\Text(110,-1)[r]{$q$}
\Text(185,25)[r]{$p$}
\Text(185,-25)[r]{$r$}

\epi
\label{pertexp}
\ee
%%%%%%%%%%%%%%%%%%%%%%%%%%%%%%%%%%%%%%%%%%%%%%%%%%%%%%%%%%%%%%
with $q+p+r=0$.
${\cal K}_{\nu\beta}$ is the conventional one-particle irreducible connected
ghost-ghost-gluon-gluon kernel appearing in the 
QCD skeleton expansion~\cite{Marciano:su,Bar-Gadda:1979cz}. Notice that 
$H_{\alpha\beta}$ appears in the 
all-order Slavnov-Taylor identity satisfied by the 
conventional three-gluon vertex~\cite{Ball:1980ax}, and
is related to the conventional gluon-ghost
vertex ${\gb}_{\beta}(p,r,q)$ 
by~\cite{Marciano:su,Bar-Gadda:1979cz,PasTar}, 
\be
q^{\alpha} H_{\alpha\beta}(p,r,q) = {\gb}_{\beta}(p,r,q)\,. 
\ee
Using the diagrammatic definition of $H_{\alpha\beta}$ shown
in Eq.(\ref{pertexp}), we may recast
Eq.(\ref{gpert2}) 
in a dressed-loop expansion
as follows, 
%%%%%%%%%%%%%%%%%%%%%%%%%%%%%%%%%%%%%%%%%%%%%%%%%%%%%%%%%%%%
\be
\bpi(0,70)(75,-25)

\PhotonArc(10,0)(25,0,180){-1.5}{8.5}
\DashCArc(10,0)(25,180,360){1}
\DashArrowLine(-10.8,-13.7)(-11.8,-12.3){1}
\DashArrowLine(27.2,-17.8)(26.2,-18.8){1}

\PhotonArc(140,0)(25,0,180){-1.5}{8.5}
\DashCArc(140,0)(25,180,360){1}
\PhotonArc(190,0)(25,0,180){-1.5}{8.5}
\DashCArc(190,0)(25,180,360){1}
\DashArrowLine(119.2,-13.7)(118.2,-12.3){1}
\DashArrowLine(157.2,-17.8)(156.2,-18.8){1}
\DashArrowLine(211.7,-12.3)(210.8,-13.8){1}
\DashArrowLine(173.6,-18.8)(172.6,-17.8){1}
\Vertex(115,-0.75){1.8}
\Vertex(215,-0.75){1.8}

\GCirc(190,23.5){12}{0.8}
\GCirc(190,-25){12}{0.8}
\GCirc(165,0){12}{0.8}
\GCirc(140,23.5){12}{0.8}
\GCirc(140,-25){12}{0.8}
\GCirc(10,23.5){12}{0.8}
\GCirc(10,-25){12}{0.8}
\Vertex(-15,-0.75){1.8}
\Vertex(35,-0.75){1.8}

\Text(166,0)[c]{$\scriptstyle{{\cal K}_{\nu\rho}}$}
\Text(140,-25)[c]{$\scriptstyle{D}$}
\Text(140,23.5)[c]{$\scriptstyle{\Delta_{\mu\nu}}$}
\Text(190,-25)[c]{$\scriptstyle{D}$}
\Text(190,23.5)[c]{$\scriptstyle{\Delta_{\rho\sigma}}$}
\Text(109,-0.75)[c]{$\alpha$}
\Text(221.5,-0.75)[c]{$\beta$}
\Text(-22,-0.75)[c]{$\alpha$}
\Text(40.0,-0.75)[c]{$\beta$}
\Text(10,-25)[c]{$\scriptstyle{D}$}
\Text(10,23.5)[c]{$\scriptstyle{\Delta_{\mu\nu}}$}
\Text(-45,-1)[r]{$\Lambda_{\alpha \beta}(q)\,\,=$}
\Text(75.0,-0.75)[c]{$+$}
\epi
\label{Lab}
\ee
%%%%%%%%%%%%%%%%%%%%%%%%%%%%%%%%%%%%%%%%%%%%%%%%%%%%%%%%%%%%%%%%%%%%

\medskip

\medskip

\medskip
Due to the transversality of both $\widehat{\Pi}_{\mu\nu}(q)$ and 
$\Pi_{\mu\nu}(q)$, if we 
write 
\be
\Lambda_{\alpha \beta}(q)= g_{\alpha\beta} G(q^2) + 
q_{\alpha}q_{\beta} L(q^2),
\ee
it turns out that 
the fundamental all-order BQI between    
$\Delta(q)$ and $\widehat{\Delta}(q)$ involves only $G(q^2)$, and is given by \cite{Gambino:1999ai,Binosi:2002ez}
\be
\widehat{\Delta}(q) = \left[1+G(q)\right]^2 \, \Delta (q)\,.
\ee
It is then elementary to demonstrate that the full propagators (in the Feynman gauge) are related by 
\be
\Delta_{\mu\nu}(q) = 
\left[1+G(q)\right]^2 \widehat{\Delta}_{\mu\nu}(q)  - i 
G(q^2)\, \left[1+G(q)\right]\, \frac{q_{\mu}q_{\nu}}{q^4} \,. 
\label{bvsqtwopfc}
\ee

The process of replacing  $\Delta_{\mu\nu} \to \widehat{\Delta}_{\mu\nu}$ will
therefore introduce the function $G$ inside the loops; however,
the theoretical and practical consequences of this operation are not clear to us at this point.
Some of the various possibilities that one might  envisage include: 
to study the dynamical equation for  $G$ ({\it viz.} Eq.~(\ref{gpert2}))
together with the SD of $\widehat{\Delta}$, as a coupled system~\cite{Grassi:2004yq};
attempt to 
reabsorb the $G$'s into a redefinition of the vertices appearing in the diagrams, together with their
corresponding SD equations; consider the diagrams containing 
$G's$ as being of higher order in the dressed loop expansion, due to the additional explicit integration 
appearing in their definition, Eq.~(\ref{gpert2}).

In the rest of this article
we will adopt what appears to be the lowest order approximation 
in this context,  setting
$\Delta_{\mu\nu} = \widehat{\Delta}_{\mu\nu}$ in the SD equation, and 
$G=0$ everywhere else.

%%%%%%%%%%%%%%%%%%%%%%%%%%%%%%%%%%%%%%%%%%%%%%%%%%%%%%%%%%%%%%%%%%

\setcounter{equation}{0}
\section{General considerations for IR finite solutions}
\label{Sect:GenCon}

In this section we will briefly review some of the main 
issues involved when trying to obtain 
from the corresponding SD equation
IR finite solutions
for the gluon self-energy $\widehat{\Delta}$. 
By ``IR finite'' we mean solutions  for which 
$\widehat{\Delta}^{-1}(0)\neq 0$, which is equivalent
to saying that $\widehat{\Delta}(0)$ is finite.

\subsection{Two necessary conditions}
 
Let us start by considering two necessary 
conditions for obtaining IR finite solutions.

(i) From the one-loop dressed SD equation for the gluon self-energy 
(see Fig.~\ref{group_a})
it is clear that, since there can be at most two full gluon self-energies inside 
the diagrams, on dimensional grounds 
the value of $\widehat{\Delta}^{-1}(0)$ will in general 
be proportional to two types of seagull-like contributions,
\bea
{\cal T}_{0} &=& \int\!\,[dk]\,\widehat\Delta(k) \,,\nonumber\\
 {\cal T}_{1} &=& \int\!\,[dk]\,k^2 \, \widehat\Delta^{2}(k) \,.
\label{seagullgen}
\eea
Perturbatively,  
both  ${\cal T}_{0}$ and ${\cal T}_{1}$ 
vanish by virtue of the dimensional
regularization result 
\be
\int \!\! \frac{[dk]}{k^2} \ln^{N} (k^2)  =0 \,, \quad N =0, 1,2,\ldots
\label{dim_reg}
\ee
which guarantees the masslessness of the gluon to all orders in perturbation
theory. 
In order to permit IR finite solutions 
one must assume that
seagull-like
contributions, such as those shown in (\ref{seagullgen}), 
do not vanish non-perturbatively.
Of course, once assumed non-vanishing, both ${\cal T}_{0}$ and ${\cal T}_{1}$
are quadratically divergent, and, in order to make sense out of them, a  
suitable regularization  must be employed.

(ii) The form of the full three-gluon vertex
is instrumental for the generation
of IR finite solutions.
Specifically, as is well known from the classic papers 
on dynamical mass generation~\cite{Jackiw:1973tr},
in order to obtain 
$\widehat{\Delta}^{-1}(0)\neq 0$
 one must introduce massless poles in the full three-gluon vertex
${\g}_{\nu\alpha\beta}$, appearing in the expression for 
$\widehat{\Pi}_{\mu\nu}^{(\bf{a_1})}(q)$, Eq.(\ref{groupa})
~\cite{Baker:1980gf}.
Notice in particular that, 
whereas after allowing for non-vanishing seagull contributions
the  inclusion of 
graph $(\bf{a_2})$ is essential
for the transversality of $\widehat{\Pi}_{\mu\nu}$,
its presence does not lead to $\widehat{\Delta}^{-1}(0)\neq 0$.
Thus, if the full three-gluon vertex $\g$ 
satisfies the WI of (\ref{3gl}), but does not contain 
poles, then the seagull contribution ${\cal T}_{0}\neq 0$ of graph $({\bf a_2})$
will cancel exactly against 
analogous contributions contained in graph $({\bf a_1})$,
forcing $\widehat{\Delta}^{-1}(0)= 0$.
Put in a different way, the non-vanishing seagull contribution that 
will determine the value of $\widehat{\Delta}^{-1}(0)$ is {\it not} 
the one coming from graph $({\bf a_2})$.
This is even more evident in the non-linear 
treatment, where the term ${\cal T}_{1}$ makes its appearance; clearly, 
such a term cannot be possibly obtained from $({\bf a_2})$.

To appreciate the delicate interplay 
between the points mentioned above,
let us consider the ghostless, 
one-loop dressed version of the SD equation, by keeping only
the two graphs of group (a).
The SD equation has the general form,
\be
\widehat{\Delta}^{-1}(q^2) {\rm P}_{\mu\nu}(q)
= q^2  {\rm P}_{\mu\nu}(q) + i \left[ \widehat{\Pi}_{\mu\nu}^{(\bf{a_1})}(q) + 
 \widehat{\Pi}_{\mu\nu}^{(\bf{a_2})} \right]\,,
\label{SD1}
\ee
with 
\bea
\widehat{\Pi}_{\mu\nu}^{(\bf{a_1})}(q) &=& 
\frac{1}{2}\, C_A \, g^2 \, \int\!\! [dk]
\widetilde{\Gamma}_{\mu\alpha\beta}
\widehat{\Delta}^{\alpha\alpha'}(k) 
{\g}_{\nu\alpha'\beta'} 
\widehat{\Delta}^{\beta\beta'}(k+q)\,,
\nonumber\\
\widehat{\Pi}_{\mu\nu}^{(\bf{a_2})}  &=&  - C_A \, g^2 \,g_{\mu\nu} {\cal T}_{0}\,.
\label{polar}
\eea

Let us   
write $i\widehat{\Pi}_{\mu\nu}^{(\bf{a_1})}(q)$ in the general form
\be
i\widehat{\Pi}_{\mu\nu}^{(\bf{a_1})}(q)= 
q^2 A(q^2) g_{\mu\nu} + B(q^2) q_{\mu} q_{\nu}\,,
\label{genform}
\ee
where $A(q^2)$ and $B(q^2)$ are arbitrary dimensionless functions, 
whose precise expressions depend on the details of the 
${\g}_{\nu\alpha'\beta'}$ employed.

The transversality of $i \left[ \widehat{\Pi}_{\mu\nu}^{(\bf{a_1})}(q) + 
 \widehat{\Pi}_{\mu\nu}^{(\bf{a_2})} \right]$
implies immediately the 
condition 
\be
q^2 \left[ A(q^2)+B (q^2)\right]  =  i C_A \, g^2 \,  {\cal T}_{0}\,,
\ee
and thus the sum of the two graphs reads  
\be
i \left[ \widehat{\Pi}_{\mu\nu}^{(\bf{a_1})}(q) + 
 \widehat{\Pi}_{\mu\nu}^{(\bf{a_2})} \right]
= - q^2 B(q^2) P_{\mu\nu}(q) \,.
\ee
Clearly,
\be
\widehat\Delta^{-1}(0) = {\displaystyle\lim_{q^2\to 0}} (- q^2 B(q^2)) \, 
= {\displaystyle\lim_{q^2\to 0}} ( q^2 A(q^2))
-  i C_A \, g^2 \, {\cal T}_{0} \,.
\label{eqzero}
\ee
 
Interestingly enough, 
once the transversality of $\widehat{\Pi}_{\mu\nu}$ has been
enforced,
 the behavior of $\widehat{\Delta}^{-1}(q^2)$ is 
determined {\it solely} by $B(q^2)$. 
In particular, the value of $\widehat{\Delta}^{-1}(0)$ 
is given by  
${\displaystyle\lim_{q^2\to 0}}\left(- q^2 B(q^2)\right)$. 
Evidently, if $B(q^2)$ does not contain $(1/q^2)$ terms, one has that 
${\displaystyle\lim_{q^2\to 0}} (- q^2 B(q^2))= 0$, and therefore
$\widehat{\Delta}^{-1}(0)=0$, despite the fact that  
${\cal T}_{0}$ has been assumed to be non-vanishing.
Actually, as we
will see in the context of the linearized SD equation  
that we will study in the next section, if  ${\g}_{\nu\alpha\beta}$
does not contain massless poles, it is precisely
this latter situation that is realized, by virtue of an 
identity relating the various integrals involved.
On the other hand, if $B(q^2)$ contains $(1/q^2)$ terms, 
${\displaystyle\lim_{q^2\to 0}} (- q^2 B(q^2))\neq 0$, 
allowing for $\widehat{\Delta}^{-1}(0)\neq 0$. 
In fact, for physically acceptable solutions, one must demand that $\widehat\Delta^{-1}(0) >0$,
which imposes further restrictions on the possible  
forms of $\g$. Of course, this is not to say that the presence of poles in the vertex is 
sufficient for obtaining IR-finite solutions, because $B(q^2)$ may end up being non-singular
due to accidental algebraic cancellations.  
As we will see in a concrete example in the 
next section, the net contributions of pole terms originating from different Lorentz
structures within the same vertex may lead to an equation that does not generate mass.

The quantities $A(q^2)$ and $B(q^2)$ appearing in (\ref{genform}), 
will be 
functionals of the unknown quantity $\widehat{\Delta}$, 
as dictated by the SD equation;
their 
specific form will depend
on the details of the vertex Ansatz chosen. 
In general, 
their value at $q^2 =0$ will be linear combinations of
the two terms defined in (\ref{seagullgen}), namely
\be
\widehat\Delta^{-1}(0) = g^2 \left( a_{0} {\cal T}_{0} + a_{1} {\cal T}_{1}\right)\,,
\label{D0qd}
\ee
where the values of the numerical coefficients $a_{0}$ and  $a_{1}$ 
depend on the details of the problem.

Then, the corresponding SD equation  will read schematically
\begin{equation}
\widehat{\Delta}^{-1}(q^2) =\widehat{\Delta}^{-1}(0) + g^2 q^2 
\int\! [dk]\,\,{\cal F}\{\widehat{\Delta};q,k\} \,, 
\label{SDsch}
\end{equation}
where the functional ${\cal F}\{\widehat{\Delta};q,k\}$ is regular at 
 $q^2 =0$. Then, after renormalizing,
we will be looking for solutions of (\ref{SDsch}) that {\it simultaneously}:
({\bf a}) 
reproduce correctly the asymptotic behavior for $\widehat{\Delta}(q^2)$
predicted by the RG, {\it viz.} Eqs.~(\ref{rightRG}) and (\ref{effch});
({\bf b}) are finite at $q^2=0$; 
({\bf c}) satisfy the (appropriately regulated) constraint
of (\ref{D0qd})~\cite{foot1}. 
Turns out that points ({\bf a}), ({\bf b}), and ({\bf c})  are deeply intertwined,
in a way that we will sketch below, and further elaborate upon in Sec.~\ref{Sect:linear_toy}.

\subsection{RG behavior and the SD equation}

The asymptotic behavior that $\widehat{\Delta}(q^2)$ must satisfy in the deep UV
is given by Eqs.~(\ref{rightRG}) and (\ref{effch}).
In practice, however, it is highly non-trivial to obtain, 
from the corresponding SD equation, solutions  
displaying this asymptotic behavior. 
This difficulty is intimately connected to the approximations 
used for the (all-order) vertex  $\g$. Of course, within a full 
SD equation treatment, $\g$ satisfies its own non-linear integral equation,
which determines its structure.
One must deal then with a very complex system of coupled integral equations 
involving  $\widehat\Delta$,  $\g$, and several many-particle kernels.
The usual way to reduce the difficulty of this problem
is to resort to the gauge-technique, 
namely express  $\g$ as a functional of $\widehat\Delta$, in such a way 
as to satisfy (by construction) the first WI of Eq.~(\ref{3gl}) exactly.
This procedure fixes the ``longitudinal'' part of the vertex, 
but leaves its ``transverse'' (identically conserved) part undetermined. 
This ambiguity, in turn, leads to the mishandling of overlapping 
divergences, which manifests itself in the fact that  
(i) one cannot renormalize multiplicatively, 
but only subtractively, and 
(ii) the RG-behavior of the solutions is distorted.
In particular, one obtains solutions which asymptotically behave like  
\be
\widehat\Delta^{-1}(q^2) = 
q^2 \left[1 + w b g^2\ln\left(\frac{q^2}{\mu^2}\right)\right]^{1/w}\,,
\label{genwrrg}
\ee
with $w>1$. These solutions reproduce upon expansion the  expected 
(one-loop) perturbative result,
but non-perturbatively they miss the correct RG behavior of (\ref{rightRG}).

The first-principle remedy of the situation would require the 
full treatment of the conserved part of the vertex, in a way similar to that
followed in \cite{King:1982mk}  
for the vertex appearing in the electron and quark gap equations. 
Unfortunately, extending their method to the case of the three-gluon vertex 
is technically very involved, 
and is at the moment beyond our powers.
Instead, we propose to model the RG behavior according to the
simple prescription put forth in \cite{Cornwall:1982zr,Cornwall:1985bg}.
The basic observation is that 
the correct RGI may be restored 
if every  $\widehat\Delta(z)$ appearing inside  
${\cal F}\{\widehat{\Delta};q,k\}$
were to be multiplied (``by hand'') by a factor (see Eq.(\ref{effch}))
\be
1+ b g^2\ln(z/\mu^2) =\frac{g^2} {\overline{g}^{2}(z)}\,.
\label{gratio}
\ee 
Thus, one is effectively switching from (\ref{SDsch}) 
to the corresponding``RG-improved'' equation 
\begin{equation}
\widehat{\Delta}^{-1}(q^2) =\widehat{\Delta}^{-1}(0) + g^2 q^2 
\int\! [dk]\,\, {\cal F}\{\widehat\Delta(z) \left(g^2/\overline{g}^{2}(z)\right);q,k\}\,,  
\label{SDschRGimp}
\end{equation}
with $z=k^2$ or $z=(k+q)^2$;
equivalently, in 
terms of the manifestly RG-invariant quantities ${\widehat d}(q^2)$ and 
$\bar{g}^2(q^2)$,  
\begin{equation}
{\widehat d}^{\,-1}(q^2) ={\widehat d}^{\,-1}(0) +  q^2 
\int\! [dk]\,\, {\cal F}\{{\widehat d}(z)/\overline{g}^{2}(z) ;q,k\}  
\label{SDschd}
\end{equation}
and 
\be
{\widehat d}^{\,-1}(0) = a_0 \overline{{\cal T}}_{0} +  a_1 \overline{{\cal T}}_{1}\,,
\label{D0qdd}
\ee
where (in Euclidean space), by virtue of (\ref{ddef1})
\bea
\overline{{\cal T}}_{0} &=&  
\int\!\, [dk] \frac{{\widehat d}(k^2)}{\overline{g}^{2}(k^2)} \,,
\nonumber\\
\overline{{\cal T}}_{1} &=&  \int\!\,[dk]\, 
\frac{ k^2 \,{\widehat d}^{\,2}(k^2)}{\overline{g}^{4}(k^2)}\,.
\label{seagullgen2}
\eea

\subsection{Regularization of seagull-like terms}

Returning  the issue of the regulation of the constraint (\ref{D0qdd}), notice 
that the integrals in (\ref{seagullgen}) differ 
from those of (\ref{seagullgen2}) in an important way.
Roughly speaking, the presence of the RG logarithms in their numerators,
(contained in $\overline{g}^{2}(k^2)$ and $\overline{g}^{4}(k^2)$, respectively)  
compensates the logarithms contained in ${\widehat d}(k^2)$ and ${\widehat d}^2(k^2)$,
allowing one to regularize them 
simply by subtracting 
a unique (vanishing) integral, that of Eq.(\ref{dim_reg}) for $N=0$, 
{\it provided} the solutions satisfy certain generic conditions.

To study this in detail, let us for the moment 
concentrate on (potential) solutions of the SD equation that are qualitatively
of the general form  
\be
{\widehat d}(q^2) = \frac{\gnp(q^2)}{q^2 + m^2(q^2)}\,,
\label{ddef}
\ee
where 
\be
\gnp(q^2) = \bigg[b\ln\left(\frac{q^2 + f(q^2, m^2(q^2))}{\Lambda^2}\right)\bigg]^{-1}\,.
\label{GNP}
\ee
The function $m^2(q^2)$ may be interpreted as 
a momentum dependent ``mass'' 
with the property (to be imposed self-consistently) that $m^2(0)>0$.
In addition, we expect that $m^2(q^2)$ is a monotonically decreasing function of $q^2$, with
$m^2(q^2)\rightarrow 0$ as $q^2/\Lambda^2\to \infty$.
The quantity $\gnp(q^2)$
represents a non-perturbative version of the 
RG-invariant effective charge of QCD, going over to $\overline{g}^{2}(q^2)$ in the deep UV.
The (dimensionfull) function $f(q^2, m^2(q^2))$ is expected to be such that 
$\gnp(q^2)$ will be a monotonically decreasing function of $q^2$, with 
$f(0, m^2(0)) > \Lambda^2$. The dimensionality 
of $f$ is to be saturated by $m^2(q^2)$; thus
if one were to set 
$m^2(q^2)=0$ then one should have  $f(q^2,0) =0$. 
The presence of a $f(q^2, m^2(q^2))$ with such properties 
in the logarithm of $\gnp(q^2)$ eliminates the Landau pole, and
leads in the deep IR 
to the characteristic property of ``freezing''. 
For the analysis that follows, note also that 
\be
\frac{1}{\gnp(q^2)} - \frac{1}{\overline{g}^{2}(q^2)} = b 
\ln\left(1 + \frac{f(q^2, m^2(q^2))}{q^2}\right)\,.
\ee
It turns out that, for the proposed regularization to work, 
both $m^2(k^2)$ and $f(q^2, m^2(q^2))$ must drop ``sufficiently fast'' 
in the deep UV.

To understand this point, 
we substitute into $\overline{{\cal T}}_{0}$ of Eq.(\ref{seagullgen2})  a 
solution of the form (\ref{ddef}),
\be
\overline{{\cal T}}_{0} =  
\int\!  [dk] \frac{\gnp(k^2)}{[k^2 + m^2(k^2)]\overline{g}^{2}(k^2)}\,,
\label{tad1}
\ee
and consider $\overline{{\cal T}}_{0}^{\,{{\rm reg}}}$, 
obtained after subtracting  $\int\!\,[dk]/k^2 =0$ from $\overline{{\cal T}}_{0}$,
\bea
\overline{{\cal T}}_{0}^{\,{{\rm reg}}} &\equiv&  
\int\! [dk] \bigg(\,\frac{\gnp(k^2)}{[k^2 + m^2(k^2)]\overline{g}^{2}(k^2)} -\frac{1}{k^2}\bigg)
\nonumber\\
&=& - \int\! [dk] \frac{m^2(k^2)}{k^2\, [k^2 + m^2(k^2)]}\,
- b \int\! [dk]\, {\widehat d}(k^2)\,
\ln\left(1 + \frac{f(k^2, m^2(k^2))}{k^2}\right)\,.
\label{basreg}
\eea
Let us examine the two integrals on the RHS separately.
If $m^2(k^2)$ behaved asymptotically as 
$\ln^{-a} (k^2)$, with the anomalous mass-dimension $a >1$,
then the first integral would  converge,
by virtue of the elementary result 
\be
\int  \frac{dz}{z\, (\ln z)^{1+\gamma}} = - \frac{1}{\gamma \, (\ln z)^{\gamma}}\,.
\label{elint}
\ee
Whether such a behavior of $m^2(k^2)$ is realized or not  
must be verified directly from the 
corresponding SD equation. For example, this was 
 indeed the case for the equations studied in \cite{Cornwall:1982zr,Cornwall:1985bg},
(with  $a= 12/11$), and we will observe it again in the next section.
In fact, a faster 
asymptotic behavior of the form $\ln^{a}(k^2)/k^2$
may be obtained from non-linear versions of the 
SD equation~\cite{Cornwall:1985bg}. 
The second integral will converge as well, provided that 
$f(k^2, m^2(k^2))$ drops asymptotically at least as fast as 
$\ln^{-c} (k^2)$, with $c>0$. If, for example, $f=\rho m^2(k^2)$ (with $a >1$,
for the first integral to converge),
then the convergence condition for the second integral is automatically fulfilled.  
Notice that perturbatively $\overline{{\cal T}}_{0}^{\,{{\rm reg}}}$ 
vanishes; this is because   $m^2(k^2)=0$ to all orders, 
and therefore, since in that case also $f=0$,
both integrals on the RHS of (\ref{basreg}) vanish.

Assuming that  $m^2(k^2)$ and $f(k^2, m^2(k^2))$ behave as described above, 
then it is straightforward to verify that the difference 
$\overline{{\cal T}}_{1} - \overline{{\cal T}}_{0}$ is automatically finite. Indeed, 
applying the elementary identity $k^2 = [k^2+m^2(k^2)]-m^2(k^2)$ 
in the numerator of $\overline{{\cal T}}_{1}$, we arrive at
\be
\overline{{\cal T}}_{1} - \overline{{\cal T}}_{0} = - 
\int\!\,[dk]\,\frac{m^2(k^2) \,{\widehat d}^{\,2}(k^2)}{\overline{g}^{4}(k^2)} \,
-  b \int\! [dk]\, \frac{\gnp(k^2)}{\overline{g}^{2}(k^2)}
{\widehat d}(k^2)
\ln\left(1 + \frac{f(k^2, m^2(k^2))}{k^2}\right)\,,
\label{T0T1}
\ee
where both integrals on the RHS converge, without any additional assumptions.

%%%%%%%%%%%%%%%%%%%%%%%%%%%%%%%%%%%%%%%%%%%%%%%%%%%%%%%

Thus, the RHS of (\ref{D0qdd}) can be written as 
\be
{\widehat d}^{\,-1}(0) = (a_0 +  a_1 ) \overline{{\cal T}}_{0} +  
a_1 (\overline{{\cal T}}_{1}-\overline{{\cal T}}_{0})\,.
\label{D0qdd2}
\ee
Clearly, if we happened to have that $a_1=-a_0$, 
the RHS of (\ref{D0qdd2}) would be automatically convergent, without further need of regularization.
If $a_1 \neq -a_0$, we will replace on the RHS of (\ref{D0qdd2})
$\overline{{\cal T}}_{0})$ by $\overline{{\cal T}}_{0}^{\,reg}$, arriving at the 
regularized version of (\ref{D0qdd}) 
\be
{\widehat d}^{\,-1}_{{{\rm reg}}}(0) = (a_0 +  a_1 ) \overline{{\cal T}}_{0}^{\,{{\rm reg}}} 
+ a_1 (\overline{{\cal T}}_{1}-\overline{{\cal T}}_{0})\,.
\label{basreg1}
\ee

It is important to emphasize that the form of the solutions assumed in (\ref{ddef})
is meant to quantify the UV behavior necessary for the proposed regularization to 
work, but does not restrict their deep IR behavior.
Specifically, 
let us assume that a solution of the type 
(\ref{ddef}), to be denoted by ${\widehat d}_{\rm \,C}(q^2)$,  
satisfies the necessary asymptotic conditions, and consider a new function, 
${\widehat d}_{\rm \,N}(q^2)= {\widehat d}_{\rm \,C}(q^2) + h(q^2)$,
where $h(q^2)$
vanishes faster than ${\widehat d}_{\rm \,C}(q^2)$
in the UV, but is not restricted in the IR. 
Then, if ${\widehat d}_{\rm \,N}(q^2)$ is inserted into (\ref{basreg}), the 
corresponding integrals are still finite. 
 As we will see in Sec.~\ref{Sect:NumAn},
this situation does in fact occur: when solving the corresponding SD equation, 
in addition to
the ``canonical'' solution of the type (\ref{ddef}), we find solutions that in the UV go 
over to  (\ref{ddef}), but in the deep IR display a much sharper increase. The point is that 
these new solutions can be regulated following the same procedure outlined here.

Several comments are now in order.

(i)  The above  method for  regulating the  seagull-like  terms relies
solely  on the  integration rules  of the  only  known gauge-invariant
regularization  scheme,  namely  dimensional regularization,  together
with  the requirement of  an appropriate  momentum dependence  for the
dynamical mass.   In that sense it is  conceptually rather economical,
evoking a minimum amount of additional theoretical input.

(ii) The implementation of the proposed regularization  hinges crucially 
on the requirement that, within the given truncation scheme,
the RGI behavior can be 
encoded faithfully into the SD equation, 
and the logarithmic terms are correctly accounted for.  
In particular, the compensation (in the UV) of the 
RG logarithm contained in ${\widehat d}(k^2)$ by  the logarithm of $\overline{g}^{-2}(k^2)$,
is essential for the consistency of the ensuing regularization of (\ref{seagullgen2}).
In the absence of the compensating logarithm one would have to subtract instead 
a term $1/k^2 \ln k^2$ in order to achieve UV convergence.
The rules of dimensional regularization allow such a possibility; by virtue of the 
more general result~\cite{Collins}
\be
\int \!\! [dk] (k^2)^{\alpha} =0 \,,
\ee
valid for any value of $\alpha$, together with the elementary identity~\cite{Cornwall:1977sh}
\be
\bigg[1+bg^2 \ln (k^2/\mu^2)\bigg]^{-R} = \frac{1}{\Gamma(R)}\int_{0}^{\infty}
dt\, e^{-t} \, t^{R-1}\, (k^2/\mu^2)^{-t\, b \,g^2} 
\ee
(valid for $R>0$), one may set
\be 
\int \frac{[dk]}{k^2 [1+bg^2 \ln (k^2/\mu^2)]} =
\int_{0}^{\infty} dt\, e^{-t}\, (\mu^2)^{t b g^2} \int \!\! [dk] \,(k^2)^{-(1+t b g^2)} = 0.
\ee
Subracting such a term would eventually regulate the initial integral in the UV.
The problem, however, is in the IR: the logarithm in the denominator of the 
regulated integral would give rise to 
the very pathology one has set out to cure in the first place,
namely the Landau pole.

(iii) Of course, ${\widehat d}^{\,-1}(0)$ must be positive definite (in Euclidean space).
The regularization possibility offered by  (\ref{basreg})
was in fact appreciated in \cite{Cornwall:1982zr}, 
but was not pursued further, 
on the grounds of furnishing the ``wrong'' sign for ${\widehat d}^{\,-1}(0)$.
In the context of that work
this was indeed so, because the three-gluon 
vertex used (the analogue of $\g$) was 
completely fixed, being the tree-level vertex derived from the Lagrangian
of the massive gauge invariant Yang-Mills model.
After connecting 
 ${\cal T}_{0}$ to 
the finite vacuum expectation value of a composite scalar field 
creating $0^{+}$ glueball states, 
the 
replacement 
\begin{equation}
 \int\!  [dk] \,\Delta(k^2) 
\longrightarrow
 \Delta^{-1}(0)\int\! [dk]\, \Delta^2 (k^2) 
\label{reg_tad}
\end{equation}
was used instead. 
Assuming that the correct RG-behavior is captured by the corresponding SD equation, then
the integral on the RHS of (\ref{reg_tad}) converges due 
to the extra logarithms in the denominator.

In the case we consider here, the sign situation is more involved.
The  two integrals on the RHS of Eqs.~(\ref{basreg}) 
are positive definite 
(assuming that $m^2(k^2)>0$ and $f(k^2, m^2(k^2)) >0$ in the full range of momenta);
thus,  the sign of $\overline{{\cal T}}_{0}^{\,{{\rm reg}}}$ is fixed.
On the other hand, the sign of the RHS of (\ref{T0T1}) is not definite,
since the $\ln z$ appearing in the numerator of the second integral 
(contained in $\overline{g}^{-2}(k^2)$) becomes negative at $z<1$.
In addition, and perhaps more importantly,
the signs of $a_0$ and $a_1$
are not {\it a-priori} known either.  
This is so because the expression 
for  ${\widehat d}^{\,-1}(0)$ is determined not by the (fixed and known) sign of the 
seagull graph, but from a delicate combination 
of the coefficients of the pole terms contained in the 
full three-gluon vertex $\g$. 
In our opinion the issue of the sign should be settled 
within the strict confines of QCD and dimensional regularization. Thus,  
if a QCD-derived (approximate, to be sure) 
form for $\g$ were to yield (after applying the proposed  regularization) 
a negative sign for ${\widehat d}^{\,-1}(0)$, then one should 
be inclined to conclude 
that the dynamical mass generation is not realized, at least not in the context 
of the specific truncation scheme.

%%%%%%%%%%%%%%%%%%%%%%%%%%%%%%%%%%%%%%%%%%%%%%%%%%%%%%%%%%%%%%%%%%%%%%%%%%%%%%

\setcounter{equation}{0}
\section{The linearized SD equation}
\label{Sect:linear_toy}

In this section we will study in detail a linearized version
of the SD equation obtained in the one-loop dressed approximation,
omitting the ghosts.
The resulting equation 
is a close variant of the one presented
in \cite{Cornwall:1982zr}, but displays several 
distinct features, allowing us 
to address further points of interest.

\subsection{Linearizing the SD equation}

We start by considering the
two diagrams of group (a), shown in Fig.~\ref{group_a}.
Since we are working in the Feynman gauge of the renormalizable 
gauges, instead of the axial gauges used in \cite{Cornwall:1982zr},
the general form of $\widehat{\Delta}_{\mu\nu}$ 
is that of Eq.~(\ref{prop_cov}).
In order to be able to use the simplified 
Ansatz for the vertex given below in conjunction with the 
Lehmann representation, it is necessary to drop the longitudinal
parts of  $\widehat{\Delta}_{\mu\nu}$ {\it inside} the integrals,
using $\widehat{\Delta}_{\mu\nu}(k) = -i g_{\mu\nu} \widehat\Delta(k)$.
As we will see in a moment, omitting these terms does not interfere 
with the transversality of the external $\widehat{\Delta}_{\mu\nu}(q)$;
in a way it is like considering scalar QED, with massive scalars inside the 
vacuum polarization loop,
yielding a transverse photon self-energy.
  
 After dropping the longitudinal parts inside the loops, we obtain
\be
\widehat{\Pi}_{\mu\nu}(q) = 
\frac{1}{2}\, C_A \, g^2 \,
 \Bigg( \int\!  [dk]\,
\widetilde{\Gamma}_{\mu}^{\alpha\beta}
\widehat{\Delta}(k) 
{\g}_{\nu\alpha\beta} 
\widehat{\Delta}(k+q)
-\,2 \,d \, g_{\mu\nu}
\int\!  [dk] \, \widehat{\Delta}(k) \Bigg)\,,
\label{polar2}
\ee
with
\be
\widetilde\Gamma_{\mu\alpha\beta}= (2k+q)_{\mu}g_{\alpha\beta} -2q_\alpha g_{\mu\beta} + 
2 q_\beta g_{\mu\alpha}\,,
\label{PTV}
\ee
and
\be
q^{\nu} {\g}_{\nu\alpha\beta} = 
\left[\widehat{\Delta}^{-1}(k+q) - \widehat{\Delta}^{-1}(k)\right] g_{\alpha\beta}\,.
\label{WID}
\ee 
Then it is straightforward to check that $\widehat{\Pi}_{\mu\nu}(q)$ is transverse.

As in \cite{Cornwall:1982zr}, in order to linearize the 
SD equation of (\ref{polar2}), we resort to 
the Lehmann representation for the gluon propagator, setting~\cite{positivity}
\begin{equation}
\widehat\Delta (q^2) = \int \!\! d \lambda^2 \, \frac{\rho\, (\lambda^2)}{q^2 - \lambda^2 + i\epsilon}\,.
\label{lehmann}
\end{equation}
This way of writing $\widehat\Delta (q^2)$ allows for a relatively simple gauge-technique Ansatz for  
${\gt}$, which linearizes the resulting SD equation.
In particular, on the RHS of the first integral in (\ref{polar2})
one sets
\begin{equation}
\widehat\Delta (k) {\g}_{\nu\alpha\beta}
\widehat\Delta (k+q) 
= \int \!\! d \lambda^2 \, \rho\,(\lambda^2)  
 \frac{1}{k^2 - \lambda^2 + i\epsilon} \,
{\gt} \,
\frac{1}{(k+q)^2 -  \lambda^2 + i\epsilon}\,,
\label{propvert}
\end{equation}
where ${\gt}$ must be such as to satisfy the tree-level WI
\be
q^{\nu} {\gt} = \left[(k+q)^2 - k^2 \right] g_{\alpha\beta}
= \left[(k+q)^2 - \lambda^2 \right] g_{\alpha\beta} - (k^2 -\lambda^2)g_{\alpha\beta}\,.
\label{WIgt}
\ee 
Then it is straightforward to show by contracting both sides 
of (\ref{propvert}) with $q^{\nu}$, and employing (\ref{WIgt}) and 
(\ref{lehmann}),  that ${\gt}$ satisfies the all-order WI of Eq.(\ref{WID}).
Of course, 
choosing ${\gt} = \widetilde\Gamma_{\nu\alpha\beta}$ solves the WI, but 
as we will see in detail in what follows, 
due to the absence of pole terms it does
not allow for mass generation, in accordance with the discussion in the previous section.
Instead we propose the following form 
\bea
{\gt} &=&
\widetilde\Gamma_{\nu\alpha\beta}    +
    c_1 \left((2k+q)_{\nu} + \frac{q_{\nu}}{q^2}
\left[k^2 - (k+q)^2\right]\right)g_{\alpha\beta} 
\nonumber\\
&&+\left( c_3 + \frac{c_2}{2\, q^2}\left[(k+q)^2 + k^2 \right]\right)
\left( q_\beta g_{\nu\alpha} - q_\alpha g_{\nu\beta} \right)\,.
\label{vertpoles}
\eea
The essential feature  of this Ansatz is that, due to the inclusion of the 
$1/q^2$ pole term, it can give rise to IR finite solutions.
Note that the additional terms have the correct properties under 
Bose symmetry with respect to the two quantum legs.  
For our purposes the constants $c_1$, $c_2$, and $c_3$  are treated as arbitrary parameters,
offering the possibility of 
quantitatively examining the sensitivity of the solutions on the 
specific details of the form of the vertex.
Of course, in reality their value will be determined by the dynamics of the 
corresponding SD equation satisfied by the full vertex, a problem which is beyond our 
powers at present.

Let us next define the quantities
\bea
B(q^2,\lambda^2) &=& \int\!\! \frac{[dk]}
{(k^2 -\lambda^2)[(k+q)^2 -  \lambda^2 ]}\,,
\nonumber\\
T(\lambda^2) &=& \int\!\! \frac{[dk]}{k^2 -\lambda^2}\,.
\label{Def1}
\eea
Substituting 
in (\ref{polar}) the expression for  ${\g}$  
obtained from the 
combination of (\ref{propvert}) and (\ref{vertpoles}), after some straightforward 
algebra we obtain
\be
\widehat\Delta^{-1}(q^2) = q^2 + \frac{C_A \,g^2 \,i}{2 (d-1)}
\int \!\! d \lambda^2 \, \rho\, (\lambda^2) F(q^2,\lambda^2)\,,
\label{D1}
\ee
with 
\bea
F(q^2,\lambda^2) &=& (7d-8)\,q^2\, B(q^2,\lambda^2) 
+ 2d \bigg\{ 2 \lambda^2 B(q^2,\lambda^2) - (d-2) T(\lambda^2)\bigg\}
\nonumber\\
&&+
 d c_1\bigg[ - q^2 B(q^2,\lambda^2) + 4 \lambda^2 B(q^2,\lambda^2)+ 2  T(\lambda^2)\bigg]
\nonumber\\
&&+ 4 (d-1) c_2 \bigg[\lambda^2 B(q^2,\lambda^2) + T(\lambda^2)\bigg]
+  4 (d-1)\, c_3\, q^2 B(q^2,\lambda^2)\,.
\label{f1}
\eea
In order to 
study whether $\widehat\Delta^{-1}(0)\neq 0$,
we must determine the value of $F(0,\lambda^2)$. To that end, note the crucial
identity 
\be
2 \lambda^2 B(0,\lambda^2) = (d-2)\, T(\lambda^2)\,, 
\label{basid} 
\ee
which may be easily proved following the standard integration rules of  dimensional 
regularization. Thus, if we were to eliminate the pole terms
in the vertex (\ref{vertpoles})
by setting $c_1=c_2=0$, then, by virtue of (\ref{basid}), we have $F(0,\lambda^2)=0$.
Evidently, the terms proportional to $c_1$ and $c_2$ in (\ref{f1}) are non-vanishing 
at  $q^2=0$, even after the application of (\ref{basid}), thus yielding 
$\widehat\Delta^{-1}(0) \neq 0$. To see this in detail,
let us define 
\be
  {\overline B}(q^2,\lambda^2) \equiv  B(q^2,\lambda^2)-B(0,\lambda^2)\,,
\ee
and then replace everywhere in (\ref{f1})~ 
$B(q^2,\lambda^2) = {\overline B}(q^2,\lambda^2) + B(0,\lambda^2)$, 
using (\ref{basid}) to eliminate $\lambda^2 B(0,\lambda^2)$ in favor of 
$T(\lambda^2)$. This leads to 
\bea
F(q^2,\lambda^2) &=& \,q^2 \bigg\{(7d-8) - d c_1 + 4 (d-1) c_3\bigg\} 
\bigg( B(0,\lambda^2)+ {\overline B}(q^2,\lambda^2)\bigg) 
\nonumber\\
&+&  4\bigg[d(1+c_1)+(d-1)c_2 \bigg]\,\lambda^2 {\overline B}(q^2,\lambda^2)
+ 2d\,(d-1)\,(c_1+c_2)\,T(\lambda^2)\,, 
\label{f2}
\eea
from which follows immediately that 
\be
\widehat\Delta^{-1}(0) =  d\,(c_1+c_2)\, C_A \,g^2\, i
\int\! [dk] \widehat\Delta (k^2)\,.
\label{D0}
\ee 
Thus,
the value of $\widehat\Delta^{-1}(0)$ is determined solely from
the  singular part  of graph  $(a_1)$, in  agreement with  the general
discussion of the  previous section. In particular, the  
seagull term  corresponding to graph (${\bf a_2}$),
due to the
aforementioned  cancellation, imposed  by  (\ref{basid}), 
 does not enter in the 
expression for $\widehat\Delta^{-1}(0)$.
This is of course  not to say that (${\bf a_2}$)
is irrelevant; on the contrary, as we have seen, the role of (${\bf a_2}$) is crucial
in enforcing transversality, which  eventually allows one to arrive at
Eq.(\ref{eqzero}) and Eq.(\ref{D0}).
Notice also that the factor determining the value of $\widehat\Delta^{-1}(0)$ is 
the sum $c_1+c_2$; thus, one could envisage the possibility of contributions from
pole terms pertaining to different Lorentz structures canceling 
against each other, or yielding the wrong sign for $\widehat\Delta^{-1}(0)$ . 

\subsection{Further algebraic manipulations}

We will now further manipulate Eq.~(\ref{f2}). 
The term proportional to $B(0,\lambda^2)$ on the RHS of (\ref{f2}) diverges, and is to 
be absorbed into the wave-function renormalization constant, 
soon to be introduced. Using the expression
for $B(0,\lambda^2)$ obtained directly from (\ref{Def1}),
 we can write it in the alternative form 
\be
\int \!\! d \lambda^2 \, \rho\, (\lambda^2) 
\int \!\! \frac{[dk]}{(k^2 -\lambda^2)^2} 
= - \int \!\! [dk] \frac{\partial }{\partial  k^2} 
\int \!\! d \lambda^2 \, \frac{\rho\, (\lambda^2)}{k^2 -\lambda^2}
= - \int \!\! [dk] \frac{\partial \widehat\Delta(k^2)}{\partial k^2} \,,
\label{I0}
\ee
where we have assumed that the order of integration may be changed.
In addition, we will use the elementary result
\be
{\overline B}(q^2,\lambda^2) = \frac{-i}{16\pi^2} 
\int_{0}^{1} dx \ln\left(1 + \frac{q^2 x(x-1)}{\lambda^2} \right)\,,
\ee
together with the following identities ~\cite{foot2}
\bea
\int \!\! d \lambda^2 \, \rho\, (\lambda^2) 
\int_{0}^{1} dx \ln\left(1 + \frac{q^2 x(x-1)}{\lambda^2} \right)
&=& \int^{q^2/4}_{0}\!\!\!dz  \left(1-\frac{4z}{q^2}\right)^{1/2} \widehat\Delta(z)\,,
\nonumber \\
\int \!\! d \lambda^2 \, \rho\, (\lambda^2) \,\lambda^2
\int_{0}^{1} dx\, \ln\left(1 + \frac{q^2 x(x-1)}{\lambda^2} \right)
&=& 
\int^{q^2/4}_{0}\!\!\!dz \,z \,\left(1-\frac{4z}{q^2}\right)^{1/2} \widehat\Delta(z)\,,
\label{Integrals}
\eea
which allow us to rewrite the RHS  manifestly in terms of the unknown function $\widehat\Delta$.
At this point it is obvious that the perturbative result, which must be 
proportional to $(7d-8)$ [see $\widehat\Pi_{\mu\nu}^{{\bf (\widehat a)}}(q)$ in (\ref{split})], 
will be distorted by the presence of the other two terms inside the curly brackets
on the RHS of (\ref{f2}).
To avoid this we will use the freedom in choosing the value of $c_3$, and 
fix it such that $4(d-1)c_3=dc_1$. After this, using the results given above, 
setting $d=4$ everywhere except in the measure,  
and defining 
\be
\tilde b \equiv \frac{10 \,C_A}{48\pi^2}\,,\,\,\,\,\,\,\,\,
\sigma\,\equiv \, \frac{6\,(c_1+c_2)} {5}\,, \,\,\,\, \,\,\,\, \gamma\,\equiv \,\frac{4+4\,c_1+3\,c_2} {5}\,,
\label{coef}
\ee
we arrive at the integral equation
\bea
\widehat\Delta^{-1}(q^2) &=& q^2 \Bigg\{ 1 - \frac{{\tilde b} g^2 i}{\pi^2}
\int \!\! d^d k \frac{\partial \widehat\Delta(k^2)}{\partial k^2}
+ {\tilde b} \,g^2 
\int^{q^2/4}_{0}\!\!\!dz \,
\left(1-\frac{4z}{q^2}\right)^{1/2}
\widehat\Delta(z)\Bigg\}
\nonumber\\
&& + \,
\gamma {\tilde b} g^2  
\int^{q^2/4}_{0}\!\!\!dz \,z \,
\left(1-\frac{4z}{q^2}\right)^{1/2}
\widehat\Delta(z) \, \,
+ \widehat\Delta^{-1}(0) \,,
\label{sd2}
\eea
with
\be
\widehat\Delta^{-1}(0) =  \frac{i {\tilde b} g^2 \sigma}{\pi^2}
\int\! d^d k \,\widehat\Delta (k^2)\,.
\label{D0M}
\ee 

Next consider the Euclidean version of (\ref{sd2}); to that end
we set  $q^2 = -q^2_{\chic E}$, with  $q^2_{\chic E} >0$ 
the positive square of a 
Euclidean four-vector, 
define the Euclidean propagator as 
\be
\widehat\Delta_{\chic E} (q^2_{\chic E}) = - \widehat\Delta (-q^2_{\chic E}), 
\ee
and the integration measure $[dk] = i[dk]_{\chic E}= i d^d k_{\chic E}/(2\pi)^4 $. 
To avoid notational clutter, we will suppress the subscript ``E''  
everywhere except in the $d^d k$ measure. Then we have 
\bea
{\widehat\Delta }^{-1}(q^2) &=& q^2 \Bigg\{ 1+ \frac{{\tilde b} g^2 i}{\pi^2}
\int \!\! d^d k_{\chic E} \frac{\partial \widehat\Delta(k^2)}{\partial k^2}
+ {\tilde b} \,g^2 
\int^{q^2/4}_{0}\!\!\!dz \left(1-\frac{4z}{q^2}\right)^{1/2} {\widehat\Delta}(z) \Bigg\}
\nonumber\\
&& + \,
\gamma {\tilde b} g^2 
\int^{q^2/4}_{0}\!\!\!dz \,z \,
\left(1-\frac{4z}{q^2}\right)^{1/2}\widehat\Delta(z) 
+ {\widehat\Delta }^{-1}(0)\,,
\eea
where from now on $q^2$  stands for the (positive) square of a Euclidean vector, and
\be
\widehat\Delta^{-1}(0) = - \frac{{\tilde b} g^2 \sigma}{\pi^2}
\int\! d^d k_{\chic E} \,\widehat\Delta (k^2)\,.
\label{D02}
\ee

\subsection{Renormalization}
In order to renormalize the equation, 
first we define the bare and renormalized quantities 
as follows:
\be
g_{\rm o} = {\widehat Z}_{g} \,g \,,\,\,\,\,\, 
{\widehat A}_{\rm o}^{\mu} = {\widehat Z}^{1/2}_{A} \, {\widehat A}^{\mu} \,,\,\,\,\,\,
{\widehat\Delta}_{\rm o}(q) = {\widehat Z}_{A} \, {\widehat\Delta}(q)\,,
\label{rendef}
\ee
and the fundamental QED-like relation ${\widehat Z}_{g} = {\widehat Z}^{-1/2}_{A}$,
which holds in the PT-BFM framework, by virtue of the Abelian-type WIs 
satisfied.
Then, it is straightforward to verify 
that 
the net effect of renormalizing (\ref{polar2}), or subsequently (\ref{sd2}), 
is to simply multiply  its RHS 
by ${\widehat Z}_{A} $ and replace all
bare quantities by renormalized ones.

However, as usually happens at this level of approximation, 
where the overlapping divergences are not properly accounted for, 
due to the ambiguities in the longitudinal parts of $\gt$, 
one 
is forced to  renormalize subtractively instead of  multiplicatively.
This amounts to interpreting ${\widehat Z}_{A}$ as an infinite constant  
that renders the product 
\be
{\widehat Z}_{A}\left(1+ \frac{{\tilde b} g^2 i}{\pi^2}
\int \!\! d^d k_{\chic E} \frac{\partial \widehat\Delta(k^2)}{\partial k^2}\right) = K
\ee
finite, and setting ${\widehat Z}_{A}=1$ in all other terms. 
This procedure leads to
\bea
{\widehat\Delta}^{-1}(q^2) &=& q^2 \Bigg\{ K
+ {\tilde b} \,g^2 
\int^{q^2/4}_{0}\!\!\!dz \, \left(1-\frac{4z}{q^2}\right)^{1/2}\,
{\widehat\Delta}(z)\Bigg\}\nonumber\\
&+& \, \gamma {\tilde b} g^2 
\int^{q^2/4}_{0}\!\!\!dz \,z \,\left(1-\frac{4z}{q^2}\right)^{1/2}
\widehat\Delta(z) 
\,\,+ {\widehat\Delta}^{-1}(0)\,.
\label{sd4}
\eea
The renormalization constant $K$ is to be fixed by the condition 
\begin{equation}
{\widehat\Delta}^{-1}(\mu^2)=\mu^2\,,
\label{ren_cond}
\end{equation}
with  $\mu^2$ a Euclidean momentum, satisfying $\mu^2 \gg  \Lambda^2$, yielding
\be
K = 1 - {\tilde b} g^2 
\int^{\mu^2/4}_{0}\!\!\!dz \, 
\left(1+ \gamma\,\frac{z}{\mu^2}\right)\,
\left(1-\frac{4z}{\mu^2}\right)^{1/2}
{\widehat\Delta}(z)\,.
\ee

\subsection{Renormalization-group analysis}

We next study the UV behavior predicted by the integral equation 
(\ref{sd4}) for ${\widehat\Delta}(q^2)$. To begin with, notice that 
the perturbative one-loop result may be recovered 
by replacing 
on the RHS of (\ref{sd4}) ${\widehat\Delta}(z) \to 1/z$, the tree-level value; 
then, 
the second term vanishes, and 
after setting  $(1- 4z/q^2)^{1/2}\to 1$ and 
$(1- 4z/\mu^2)^{1/2}\to 1$ in the first term (curly brackets), 
we obtain
${\widehat\Delta}^{-1}(q^2)|_{pert}= q^2 \left(1+ {\tilde b} \,g^2 \ln(q^2/\mu^2)\right)$. 
However, if one were to solve this equation non-perturbatively, one would discover that, 
even though the perturbative result is correctly recovered, ${\widehat\Delta}(q^2)$
does not display the expected RG behavior, 
e.g. that of Eq.~(\ref{rightRG}), with $b \to {\tilde b}$.

The fact that 
this behavior is not captured by (\ref{sd4})
can be easily seen by writing down the simplified version of that equation, 
\begin{equation}
\widehat\Delta^{-1}(q^2) = q^2\left(1+ {\tilde b}g^2\int^{q^2}_{\mu^2}\!\!\! dz\,\widehat{\Delta}(z)  
\right)\,,
\label{uv}
\end{equation}
valid in the deep UV, 
and converting it into an equivalent differential equation. Setting 
$\widehat\Delta(q^2) = G(q^2)/q^2$, and $x=q^2$,  we obtain 
\begin{equation}
\frac{d\, G(x)}{d\, x} = - {\tilde b} g^2  \frac{G^3(x)}{x}\,,
\label{diffeq}
\end{equation}
which leads to
\begin{equation}
  \widehat\Delta^{-1}(q^2)= 
q^2\left[1+ 2 {\tilde b}g^2\ln\left(\frac{q^2}{\mu^2}\right)\right]^{1/2}\,.
\label{uv_sol}
\end{equation}
Upon expansion this expression yields ${\widehat\Delta}^{-1}(q^2)|_{pert}$ correctly,
but differs from (\ref{rightRG}). 
The fundamental 
reason for this discrepancy can be essentially traced back to having 
carried out the renormalization subractively instead of multiplicatively 
\cite{Cornwall:1982zr,Cornwall:1985bg}, thus 
distorting the RG structure of the equation.

In order to restore the correct RG behavior at the level of (\ref{sd4}),
we will use the procedure explained in Sec.~\ref{Sect:GenCon}, 
substituting in the integrands on the RHS of (\ref{sd4})
\be
\widehat\Delta(z) \longrightarrow \frac{g^2\, \widehat\Delta(z)} {\overline{g}^{2}(z)} \,.
\label{gratio1}
\ee 
Then (\ref{sd4}) may be  
rewritten in terms of two RG-invariant quantities, ${\widehat d}(q^2)$ and 
$\overline{g}^2(q^2)$, 
as follows
\bea
{\widehat d}^{\,-1}(q^2) &=& q^2 \Bigg\{ K^{\prime}
+ {\tilde b} 
\int^{q^2/4}_{0}\!\!\! dz\, \left(1-\frac{4z}{q^2}\right)^{1/2}
\frac{{\widehat d}(z)}{\overline{g}^2(z)}\Bigg\}
\nonumber\\
&+&\,
\gamma {\tilde b}
\int^{q^2/4}_{0}\!\!\!dz \,z \,
\left(1-\frac{4z}{q^2}\right)^{1/2}
\frac{{\widehat d}(z)}{\overline{g}^2(z)} \,\,
+ {\widehat d}^{\,-1}(0) \,,
\label{sd5}
\eea
with 
\be
K^{\prime} = \frac{1}{g^2} - {\tilde b}  
\int^{\mu^2/4}_{0}\!\!\!dz \, 
\left(1+\gamma\,\frac{z}{\mu^2} \right)\, 
\left(1-\frac{4z}{\mu^2}\right)^{1/2}\,
\frac{{\widehat d}(z)}{\overline{g}^2(z)}\,,
\label{constreuc}
\ee
and
\be
{\widehat d}^{\,-1}(0) = 
 -  \frac{{\tilde b}\sigma}{\pi^2}
\int\! d^d k_{\chic E} \,\frac{{\widehat d} (k^2)}{\overline{g}^2(k^2)}\,.
\label{d01}
\ee 
It is easy to see now that Eq.(\ref{sd5}) yields the correct UV behavior for 
${\widehat d}(q^2)$, given in (\ref{ddef1}). For example, converting (\ref{sd5})
into a differential equation, and setting  
${\widehat d}(q^2)= F(q^2)/q^2$, the equivalent of (\ref{diffeq}) now reads
\begin{equation}
\frac{d\, F(x)}{d\, x} = - {\tilde b} \frac{F^3(x)}{x\, \bar{g}^2(x)} \,,
\label{diffeq1}
\end{equation}
whose solution is $ F(x) = \overline{g}^2(x)$, as announced.

\subsection{Asymptotic behavior of $m^2(q^2)$}

As has been discussed  in Sec.~\ref{Sect:GenCon},
the ability to regularize condition (\ref{d01}) following the 
properties of dimensional regularization 
depends crucially on 
the asymptotic behavior of $m^2(q^2)$. 
It is therefore essential to determine the asymptotic behavior that 
Eq.(\ref{sd5}) yields for  
$m^2(q^2)$ in the deep UV. 
To that end, substitute (\ref{ddef}) on both sides of (\ref{sd5}), and 
consider the limit where $q^2$ is large. 
The integral proportional to $\gamma$ may be written as 
\be
\int^{q^2/4}_{0}\!\!\!dz 
\left(1-\frac{4z}{q^2}\right)^{1/2}
\frac{z}{z+m^2(z)} = \frac{q^2}{6} - 
\int^{q^2/4}_{0}\!\! dz \left(1-\frac{4z}{q^2}\right)^{1/2}
\frac{ m^2(z)}{z+m^2(z)} \,.
\ee
Then
setting $(1- 4z/q^2)^{1/2}\to 1$, and dropping terms that do not 
grow logarithmically with $q^2$, (\ref{sd5}) reduces to
\bea
[q^2 + m^2(q^2)]\ln \left(q^2/\Lambda^2\right)
=  q^2   
\int^{q^2/4}_{0}\! 
\frac{dz}{z+m^2(z)} 
-
\gamma
\int^{q^2/4}_{0}\!\! dz 
\frac{ m^2(z)}{z+m^2(z)}\,.
\label{sd61}
\eea
Let us then separate the terms that 
go like  $q^2 \ln q^2$ and $m^2(q^2)\ln q^2$; obviously the first integral on the RHS 
compensates the   $q^2 \ln q^2$ on the LHS. Then setting 
\be
 m^2(q^2) \sim  m^2_{0} \ln^{-a} \left(q^2/\Lambda^2\right)\,,
\label{uv_mass} 
\ee
we find that the two sides of the equation can be made equal if
\be
a= 1+ \gamma\,.
\ee
Thus, provided that $\gamma >0$, the momentum dependence of the mass term 
in the deep UV is of the type 
needed for the regularization of Eq.(\ref{basreg}) to go through.

\setcounter{equation}{0}
\section{Numerical Analysis}
\label{Sect:NumAn}

The equation we will solve is (\ref{sd5}) with the renormalization constant 
$K^{\prime}$ of (\ref{constreuc}) explicitly incorporated, e.g. 

\bea
{\widehat d}^{\,-1}(q^2) &=& q^2 \Bigg\{ \frac{1}{g^2}
+ {\tilde b}\Bigg( 
\int^{q^2/4}_{0}\! dz\, \left(1-\frac{4z}{q^2}\right)^{1/2}
\frac{{\widehat d}(z)}{\bar{g}^2(z)}
- 
\int^{\mu^2/4}_{0}\!dz \, 
\left(1-\frac{4z}{\mu^2}\right)^{1/2}\,\frac{{\widehat d}(z)}{\bar{g}^2(z)} \Bigg)\Bigg\}
\nonumber\\
&+&\,
\gamma {\tilde b}\Bigg(
\int^{q^2/4}_{0}\!dz \,z \,
\left(1-\frac{4z}{q^2}\right)^{1/2}
\frac{{\widehat d}(z)}{\bar{g}^2(z)} \,\,
- \frac{q^2}{\mu^2} \int^{\mu^2/4}_{0}\!dz  \, z  
\left(1-\frac{4z}{\mu^2}\right)^{1/2}\,\frac{{\widehat d}(z)}{\bar{g}^2(z)}\Bigg)
\nonumber\\
&+& 
{\widehat d}^{\,-1}(0) \;
\label{sd6}
\eea
with  ${\widehat d}^{\,-1}(0)$ given by (\ref{d01}), eventually to 
be replaced by its regularized expression,
according to (\ref{basreg}).
In particular, if  the numerical solution obtained for ${\widehat d}(q^2)$  
has the general form of Eq.(\ref{ddef}),  ${\widehat d}^{\,-1}(0)$ will be given by 
 \be
{\widehat d}^{\,-1}(0) = 
  {\tilde b}\sigma \left[
 \int_0^{\infty}\! dz \frac{m^2(z)}{ z + m^2(z)}\,
+ \tilde{b} \int_0^{\infty}\! dz\,z \; {\widehat d}(z)\,
\ln\left(1 + \frac{f(z, m^2(z))}{z}\right) \right]  \,,  
\label{dcorn}
\ee     
If  the solutions deviate in the IR from Eq.(\ref{ddef}),
an analogous expression can be obtained, (see below) 
in accordance with the discussion following Eq.(\ref{basreg1}). 

Equations~(\ref{sd6}) and (\ref{dcorn}) form  a system of  equations that must
 be  solved simultaneously.  The  role of  (\ref{sd6}) is  to furnish
 possible  solutions  for  ${\widehat  d}(q^2)$,  while  (\ref{dcorn})
 constrains  them or the  value of  the parameters  involved.  Roughly
 speaking, the strategy for solving the system is the following.  Note
 first  that (\ref{sd6})  contains  no information  for  the value  of
 ${\widehat    d}^{\,-1}(0)$;    at    $q^2=0$,   one    obtains    an
 identity. Therefore,  one chooses  an arbitrary value  for ${\widehat
 d}^{\,-1}(0)$, plugs it in on the RHS of (\ref{sd6}), and proceeds
 to solve the integral equation. Given such a solution, one
 must then substitute  it   in  (\ref{dcorn})  and  calculate   the  value  of
 ${\widehat d}^{\,-1}(0)$ obtained after the integration, and compare
 it  with  the  value   of  ${\widehat  d}^{\,-1}(0)$  chosen  at  the
 beginning, the objective being  to reach coincidence between the two
 values.   Assuming that  the value  of $\sigma$  is considered  to be
 fixed,  one must  then repeat  this procedure,  varying  the initial
 value of ${\widehat d}^{\,-1}(0)$,  until agreement has been reached.
 If instead  one is free to choose  the value of $\sigma$,  then for a
 given initial  value of ${\widehat d}^{\,-1}(0)$  one varies $\sigma$
until compliance has been achieved.
 In  this  article we  will  follow  the  latter philosophy,  treating
 $\sigma$ as an adjustable parameter.

Specifically, for each value of ${\widehat  d}^{-1}(0)$ chosen, 
we should vary $\gamma$
and  $\sigma$,  in   order  to  scan  the  two-parameter  space  of
solutions. In practice, we choose  to reduce the number of parameters down to one,
namely $\sigma$,
since from the ensuing  numerical analysis it becomes clear  
the  dependence  of the solution shows a very mild dependence on $\gamma$.
Therefore, we rewrite $\gamma$, given by Eq.(\ref{coef}), as
\be
\gamma=\frac{2}{3}\sigma + \frac{4}{5} -\frac{c_2}{5} \,,
\label{ng}
\ee
and then we set $c_2=0$, in the order to keep
$\sigma$ as the only free parameter. The  small differences produced
when $c_2$ is non-vanishing will be commented later on.

For the numerical treatment
we define a logarithmic grid for the variables 
$q^2$ and  $z$;  this improves the  accuracy of the algorithm in the  small $q^2$
region, since  the size of  the steps is made smaller  for IR
momenta.  We  split the grid  into  two  region:   $[0,\mu^2]$  and
$(\mu^2,\Lambda_{\rm{UV}}]$.  Such splitting  in needed for imposing on
${\widehat  d}(q^2)$   the  renormalization  condition (given  by
Eq.(\ref{ren_cond})) at a perturbative scale $\mu^2$. Typically, we chose
$\mu^2=M^2_{\rm     Z}=(91.18)^2      \;     \mbox{GeV}^{\,2}$     and
$\Lambda_{\rm{UV}}=10^{\,6} \;\mbox{GeV}^{\,2}$.  Furthermore, one has
to   specify   the   coupling   $g^2=\bar{g}^2(\mu^2)$   entering   in
Eq.(\ref{sd6}); its value is obtained from Eq.(\ref{effch}), 
where we have properly replaced $b \rightarrow {\tilde b}$, and 
used as input a  value of $\Lambda = 300 \;\mbox{MeV}$ for the QCD mass scale.
Then,  we  solve  the  integral equation iteratively,
starting out with an initial trial function, and  
using as a convergence criterion that the relative difference 
between the input and the output should be smaller than $10^{\,-8}$.

The general trend displayed by the solutions is that
the characteristic IR plateau (``freezing'') becomes increasingly narrower
as one increases the value of ${\widehat d}(0)$. 
We have found two types of solutions, depending on the initial value chosen
for ${\widehat d}^{\,-1}(0)$:
(i) For values of ${\widehat    d}^{\,-1}(0)$  within the range 
$[0.01\;\mbox{GeV}^{\,2},0.07  \;\mbox{GeV}^{\,2}] $ the solutions
can be perfectly  fitted by Eq.(\ref{ddef}); we 
will refer to them as ``canonical''.
(ii) For  ${\widehat  d}^{\,-1}(0)   <  0.01\;\mbox{GeV}^{\,2} $ the solutions
can be fitted by Eq.(\ref{ddef}) until relatively low values of $q^2$,
but deviate significantly in the deep IR, where they display a sharp
rise and a rather narrow plateau;
we will call such solutions ``non-canonical''.

\subsection{Canonical solutions}

\begin{figure}[ht]
\hspace{-1.5cm}
\includegraphics[scale=1.0]{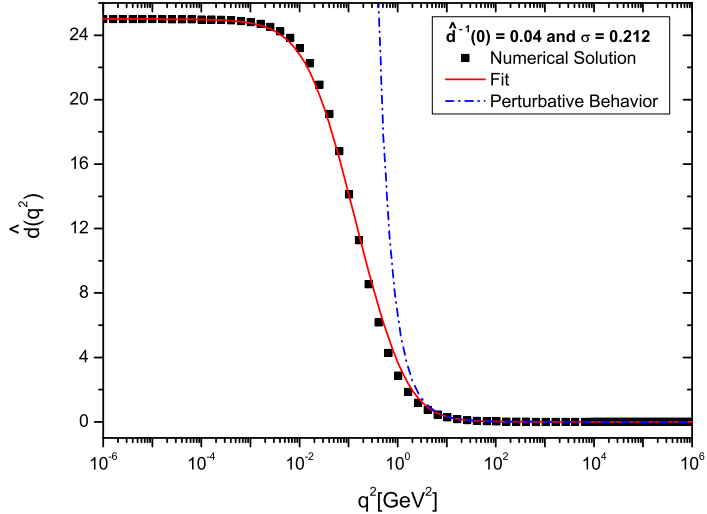}
\caption{ The black dots represent the numerical solution for 
${\widehat  d}(q^2)$ obtained for the choice 
 ${\widehat  d}^{\,-1}(0)=0.04  \;\mbox{GeV}^{\,2}$,
$\sigma=0.212$  and $\gamma=0.941$.
The continuous red
line  is   the  fit   of  Eq.(\ref{ddef}),  setting
$\rho_2=-3.208$ and $m^2_0=0.45 \, \mbox{GeV}^{\,2}$. 
These curves are
obtained  fixing   ${\widehat  d}^{\,-1}(0)=0.04  \;\mbox{GeV}^{\,2}$,
$\sigma=0.212$  and $\gamma=0.941$.  The  dashed blue line  is the 
one-loop perturbative behavior.}
\label{p1}
\end{figure}

In  Fig.~\ref{p1},  we  show   a typical case of a canonical solution
corresponding to the initial choice ${\widehat    d}^{\,-1}(0)    =   0.04
\;\mbox{GeV}^{\,2}$.  
As can  be observed
from the  plot, ${\widehat d}(q^2)$ is essentially a constant, determined by ${\widehat d}(0)$,
until the neighbourhood  of $q^2=0.01 \;\mbox{GeV}^{\,2}$; then, the  curve bends 
downward in order  to match with the perturbative  scaling behavior at a
scale of  few $\mbox{GeV}^{\,2}$.
All such solutions may be fitted very accurately by means of the ${\widehat d}(q^2)$ of Eq.(\ref{ddef}),
where the functional form of  $\gnp(q^2)$ is that of  Eq.(\ref{GNP}),
with the function $f(q^2, m^2(q^2))$ fixed as 
\be
f(q^2, m^2(q^2)) = \rho_{\,1} m^2(q^2)+ \rho_{\,2} \frac{m^4(q^2)}{q^2+m^2(q^2)} \,,
\label{func_fit}
\ee
and the dynamical mass has the form 
\be
m^2(q^2)=m^2_0\left[\ln
\left[\frac{q^2+\rho_{\,1}\,m^2_0}{\Lambda^2}\right]\Big/\ln\left[\frac{\rho_{\,1}\,m^2_0}{\Lambda^2}\right] \right]^{-a} \,,
\label{dmass}
\ee 
with exponent $a = 1+ \gamma $.
%      
%%%%%%%%%%%%%%%%%%%%%%%%Fig.6 %%%%%%%%%%%%%%%%%%%%%%%%%%%%%%%%%%%%%
%                 plot - propagator
%%%%%%%%%%%%%%%%%%%%%%%%%%%%%%%%%%%%%%%%%%%%%%%%%%%%%%%%%%%%%%%%%%%%%%
%%%%%%%%%%%%%%%%%%%%%%%%%%%%%%%%%%%%%%%%%%%%%%%%%%%%%%%%%%%%%%%%%%% 

\begin{figure}[ht]
\hspace{-1.5cm}
\includegraphics[scale=1.0]{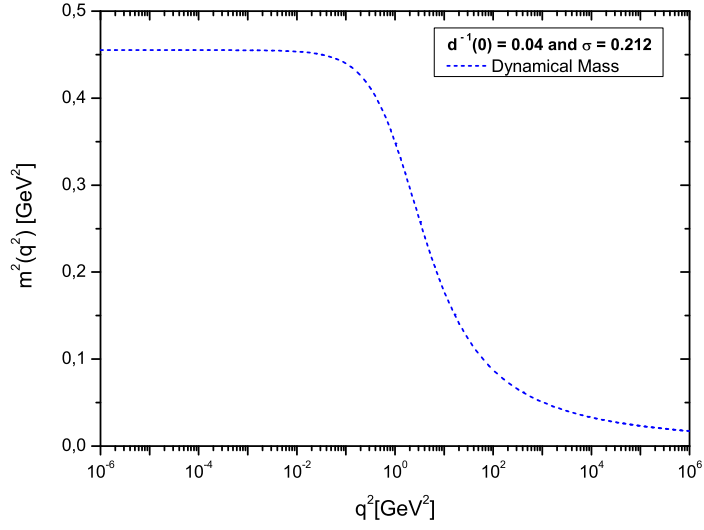}
\caption{The dynamical mass, $m^2(q^2)$,
corresponding to the solution of Fig.~\ref{p1}.}
\label{p2}
\end{figure}

\begin{figure}[ht]
\hspace{-1.5cm}
\includegraphics[scale=1.0]{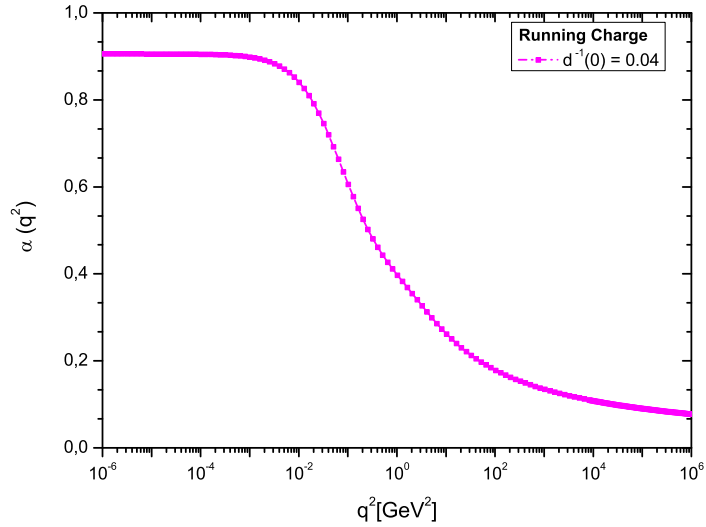}
\caption{The running charge, $\alpha(q^2)=\gnp(q^2)/4\pi$ 
corresponding to the solution of Fig.~\ref{p1}.}
\label{p3}
\end{figure}

 The dynamical  mass, $m^2(q^2)$, and
the running charge,  $\alpha(q^2)=\gnp(q^2)/4\pi$, 
 corresponding to the numerical solution presented on  the Fig.~\ref{p1},
are shown in Figs.~\ref{p2} and {\ref{p3}}, respectively. 

%%%%%%%%%%%%%%%%%%%%%%%%Fig.7 %%%%%%%%%%%%%%%%%%%%%%%%%%%%%%%%%%%%%
%          plot mass
%%%%%%%%%%%%%%%%%%%%%%%%%%%%%%%%%%%%%%%%%%%%%%%%%%%%%%%%%%%%%%%%%%%%%%
%%%%%%%%%%%%%%%%%%%%%%%%%%%%%%%%%%%%%%%%%%%%%%%%%%%%%%%%%%%%%%%%%%% 
%     
%%%%%%%%%%%%%%%%%%%%%%%%Fig.8 %%%%%%%%%%%%%%%%%%%%%%%%%%%%%%%%%%%%%
%             plot -  coupling
%%%%%%%%%%%%%%%%%%%%%%%%%%%%%%%%%%%%%%%%%%%%%%%%%%%%%%%%%%%%%%%%%%%%%%%
%%%%%%%%%%%%%%%%%%%%%%%%%%%%%%%%%%%%%%%%%%%%%%%%%%%%%%%%%%%%%%%%%%% 

In fact,
in all cases studied,
$\rho_{\,1}$ was fixed to the value $\rho_{\,1}=4$, 
which was found to  minimize the $\chi^2$ of the corresponding fits.
Therefore, the  unique free parameter is $\rho_{\,2}$,
since the value of $m^2_0$ can be written in terms of $\rho_{\,2}$,
simply by setting $q^2=0$, in Eq.(\ref{ddef}), i.e.,
\be
{\widehat d}^{\,-1}(0)=\tilde{b}\,m^2_0\ln\left(\frac{ f(0,m^2_0 )}{\Lambda^2}\right) \, ,
\label{lim_d0}
\ee
where
\be
f(0,m^2_0)= (4 + \rho_{\,2})\,m^2_0 \, .
\ee
Thus, the  value of  the IR
fixed point  of the running coupling  is determined  by the value
assumed by $f(0,m^2_0)$, since
\be
{\overline g}^{\,-2}_{{\chic {\rm NP}}}(0) = \tilde{b}\ln\left[\frac{(4 +\rho_{\,2})\,m^2_0 }{\Lambda^2}\right] \,,
\label{fita1}
\ee
Obviously, the maximum value obtained for $\gnp(0)$, is the one that
minimizes $(4  +\rho_{\,2})\,m^2_0$ and, at  same time, keeps  it bigger
than $\Lambda^2$,  in order to avoid the appearance 
of a shifted  version of the Landau pole.

Note that the 
term proportional to $\rho_{\,2}$ in (\ref{func_fit}) 
is necessary for optimizing
the fit for  
the   range   of  momenta $   q^2   \in   [0.01
\;\mbox{GeV}^{\,2},  2\;\mbox{GeV}^{\,2}]$,   i.e.  the  region  where
${\widehat d}(q^2)$ falls down rapidly. 
If  we were to set  $\rho_{\,2}=0$  in  Eq.(\ref{func_fit})  we
would recover automatically the solution proposed in \cite{Cornwall:1982zr};
however, such a  choice would not correspond to the best possible fit: 
our numerical solution requires bigger values for the coupling
$\gnp(q^2)$, and  therefore, $\rho_{\,2}$ assumes  negative values, as
can    be    observed   on    Fig.~\ref{p1},    where   found    that
$\rho_{\,2}=-3.208$.

It is important  to mention that 
other  functional  forms  for  $f(q^2,  m^2(q^2))$  were 
also tried; although some  of them could fit ${\widehat
d}(q^2)$ well, they have been discarded due to the appearance, at some scale, of
an undesired ``bump''  in the behavior of $\gnp(q^2)$.
  In others words, we have required that 
$\gnp(q^2)$ should be a {\it monotonically decreasing} function of $q^2$.

\begin{figure}[ht]
\hspace{-1.5cm}
\includegraphics[scale=1.0]{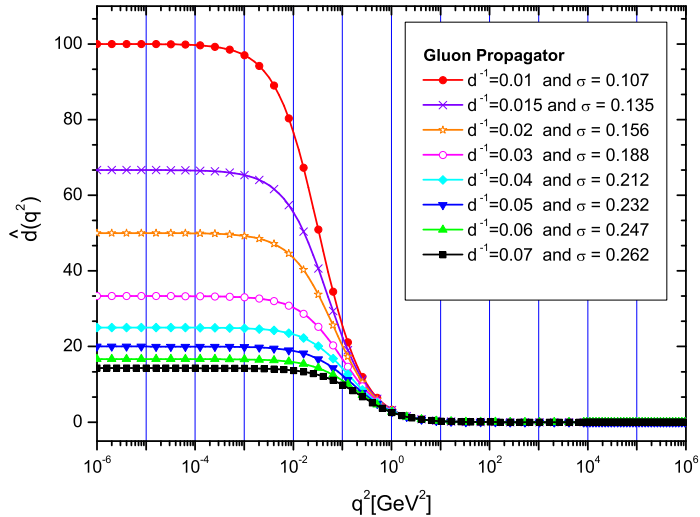}
\caption{Results for ${\widehat d}(q^2)$ for various values of ${\widehat d}^{\,-1}(0)$ (all in  $\;\mbox{GeV}^{\,2}$),
together with the respective values for $\sigma$, obtained from the constraint of Eq.(\ref{dcorn}).}
\label{p4}
\end{figure}

 In Fig.~\ref{p4} we plot a series  of
 numerical solution obtained by fixing different
 values  for ${\widehat d}^{\,-1}(0)$.  All these solutions  have been subjected  to
 the  constraint  imposed by  Eq.(\ref{dcorn}),  and their corresponding 
 values for $\sigma$ are reported in the inserted legend.
 Observe that they all behave like constants 
until   practically  the   same   scale  of $0.01   \;
\mbox{GeV}^{\,2}$ (the ``freezing'' plateau), then they 
converge down to a common value at around $1 \; \mbox{GeV}^{\,2}$, 
beyond which they start  to   develop  the known perturbative behavior.

%%%%%%%%%%%%%%%%%%%%%%%%Fig.9 %%%%%%%%%%%%%%%%%%%%%%%%%%%%%%%%%%%%%
%              plot - all propagators
%%%%%%%%%%%%%%%%%%%%%%%%%%%%%%%%%%%%%%%%%%%%%%%%%%%%%%%%%%%%%%%%%%
%%%%%%%%%%%%%%%%%%%%%%%%%%%%%%%%%%%%%%%%%%%%%%%%%%%%%%%%%%%%%%%%%%% 

\begin{figure}[ht]
\hspace{-1.5cm}
\includegraphics[scale=1.0]{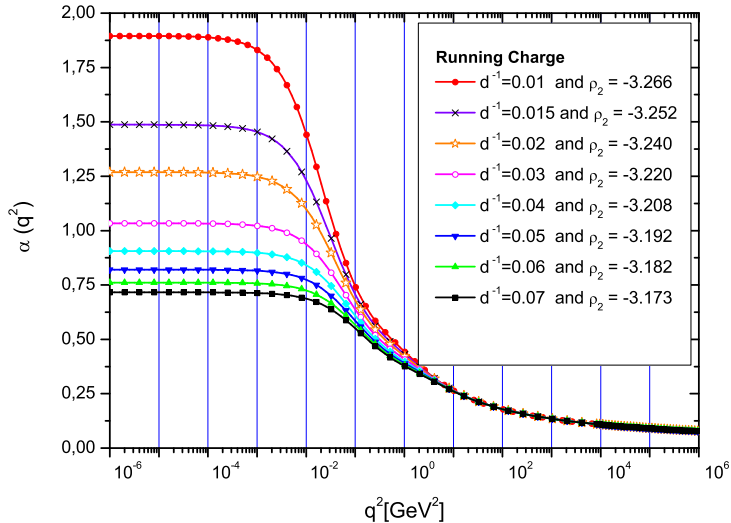}
\caption{We plot the correspondent running charge, $\alpha(q^2)$, for the gluon propagator presented on Fig.~\ref{p4}.  Clearly, we can see  as we decrease the values of ${\widehat d}^{\,-1}(0)$ the infrared fixed point, $\alpha(0)$, becomes bigger.}
\label{f5}
\end{figure}

The  running charges,  $\alpha(q^2)$, for  each solution  presented in
Fig.~\ref{p4},  are  displayed  on  Fig.~\ref{f5}. 
 Observe  that as the
value of ${\widehat d}^{\,-1}(0)$ decreases, the value of the infrared
fixed     point    of     the    running     coupling,    $\alpha(0)$,
increases. Accordingly, from Fig.~\ref{p4} and Fig.~\ref{f5}, we can
conclude that,  if smaller  values of $\sigma$  were to be  favored by
QCD,  the freezing  of  the  running coupling  would  occur at  higher
values. It should  also be noted that the  values of $\alpha(0)$ found
here  tend  to  be  slightly   more  elevated  compared  to  those  of
\cite{Cornwall:1982zr} (for the same value of $m^2_0$).

%
%%%%%%%%%%%%%%%%%%%%%%%%Fig.10 %%%%%%%%%%%%%%%%%%%%%%%%%%%%%%%%%%%%%
%               plot - all couplings
%%%%%%%%%%%%%%%%%%%%%%%%%%%%%%%%%%%%%%%%%%%%%%%%%%%%%%%%%%%%%%%%%%%%%%
%%%%%%%%%%%%%%%%%%%%%%%%%%%%%%%%%%%%%%%%%%%%%%%%%%%%%%%%%%%%%%%%%%%      
 
\begin{figure}[ht]
\hspace{-1.5cm}
\includegraphics[scale=1.0]{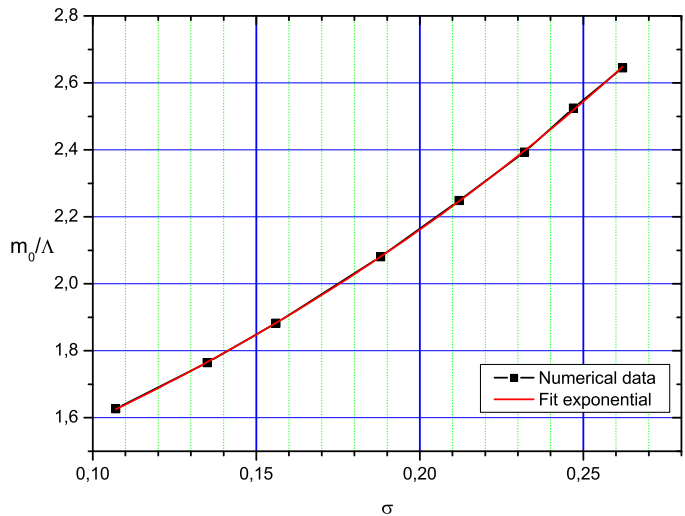}
\caption{The ratio $m_0/\Lambda$ as function of the parameter $\sigma$.
The increase is exponential, given by Eq.(\ref{fit_exp}).}
\label{ff1}
\end{figure}

In addition we analyze the dependence of the ratio $m_0/\Lambda$ on  $\sigma$,
extracted  from  Eq.(\ref{lim_d0}).
This dependence  is shown  in Fig.~\ref{ff1}, 
corresponding to the cases  presented in Fig.~\ref{p4}.
We  observe that  as we  increase the  value of
$\sigma$, namely the sum of the coefficients of 
the  massless  pole  terms  appearing  in  the three  gluon  vertex,
the  ratio   $m_0/\Lambda$  grows  exponentially as
\be
\frac{m_0}{\Lambda} = A_1\exp\left(\frac{\sigma}{t_1}\right) + y_0\,,
\label{fit_exp}
\ee
with  $A_1=0.775$, $t_1=0.25$ and $y_0=0.436$.

At this point it is natural to ask by 
 how much  these results  would change  if we were to turn on again $c_2$ in
 Eq.(\ref{ng}). We have examined values of 
$c_2$ comparable to those used for $\sigma$,
i.e. we exclude the possibility that the typically small values  
of $\sigma$ are a result of a very fine-tuned cancellation 
between large $c_1$ and $c_2$, opposite in sign.
It turns out that,  for all previous cases studied, the effect 
of including non-vanishing $c_2$'s in Eq.(\ref{ng}) is 
less than $3\%$, a fact that justifies their omission.

%%%%%%%%%%%%%%%%%%%%%%%%Fig.11 %%%%%%%%%%%%%%%%%%%%%%%%%%%%%%%%%%%%%
%                   plot - ratio versus sigma
%%%%%%%%%%%%%%%%%%%%%%%%%%%%%%%%%%%%%%%%%%%%%%%%%%%%%%%%%%%%%%%%%%%%%
%%%%%%%%%%%%%%%%%%%%%%%%%%%%%%%%%%%%%%%%%%%%%%%%%%%%%%%%%%%%%%%%%%% 
 
\subsection{Non-canonical solutions}

\begin{figure}[ht]
\hspace{-1.5cm}
\includegraphics[scale=1.0]{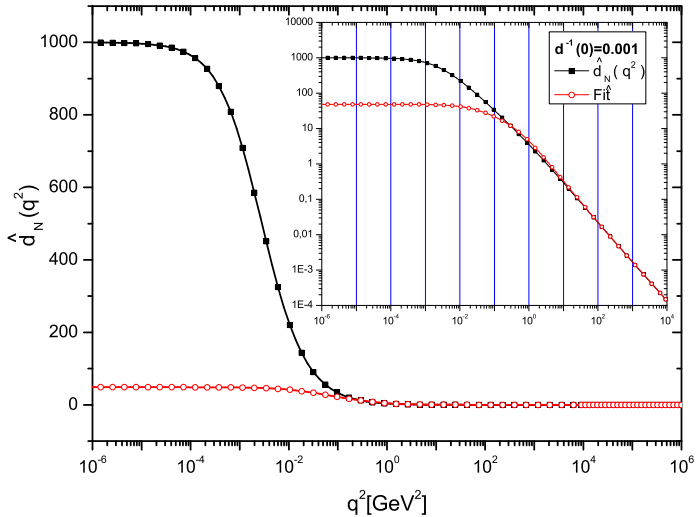}
\caption{The black square points  represent the numerical solution for
the RGI  quantity ${\widehat d}(q^2)=g^2{\widehat  \Delta}(q^2)$, when
we   fixed  ${\widehat   d}^{\,-1}(0)=0.001   \;\mbox{GeV}^{\,2}$  and
$\sigma=7.65\times 10^{\,-3}$. The red line  + circle curve is the fit
given  by  Eq.(\ref{ddef}),  setting  $\rho_2=-\pi$  and  $m^2_0=0.304
\;\mbox{GeV}^{\,2}$.  The small  plot  is the  same  graph on  log-log
scale.}
\label{dd1}
\end{figure}

As we  decrease ${\widehat d}^{\,-1}(0)$,  it becomes 
increasingly difficult to fit 
 the numerical  solution 
obtained from Eq.~(\ref{sd6}) using  the expressions given in 
(\ref{ddef}), (\ref{func_fit}), and (\ref{dmass}).
In particular, for solutions
 with ${\widehat d}^{\,-1}(0)<0.01 \; \mbox{GeV}^{\,2}$, the
 discrepancy is  relatively large,  especially  in the
 intermediate region. We will denote such non-canonical solutions by 
${\widehat d}_{\rm \,N} (q^2)$.
A typical  numerical solution
is  shown by the curve
composed by  black squares in  Fig.~\ref{dd1}. 
It is important to emphasize that, although the  form of
Eq.(\ref{ddef}) is not suited for  fitting the entire 
momentum range, one may
use  it  for  describing  a  partial  region,  i.e.  $q^2\in  [0.2  \,
\mbox{GeV}^{\,2},\Lambda_{\rm{UV}}]$, as showed by the line+circle curve on
Fig.~\ref{dd1}, where we clearly  see a quantitative agreement between
both curves. This observation is related to the discussion following 
Eq.(\ref{basreg1}), allowing us to use the same regularization 
procedure as in (\ref{dcorn}).
   
%
%%%%%%%%%%%%%%%%%%%%%%%%Fig.12 %%%%%%%%%%%%%%%%%%%%%%%%%%%%%%%%%%%%%
%                 plot - non canonical solution
%%%%%%%%%%%%%%%%%%%%%%%%%%%%%%%%%%%%%%%%%%%%%%%%%%%%%%%%%%%%%%%%%%%%%
%%%%%%%%%%%%%%%%%%%%%%%%%%%%%%%%%%%%%%%%%%%%%%%%%%%%%%%%%%%%%%%%%%%   

To see this, let us denote by ${\widehat d}_{\rm \,C} (z)$ the canonical solution that
provides the best fit
to a given ${\widehat d}_{\rm \,N} (z)$; then 
write $  {\widehat
   d}_{\rm \,N} (z) = [{\widehat d}_{\rm \,N} (z) -{\widehat d}_{\rm \,C} (z)] +
   {\widehat  d}_{\rm \,C} (z)$, and substitute into Eq.(\ref{dcorn}), to obtain  
\be
{\widehat d}^{\,-1}(0) =  {\tilde b}\sigma \bigg[ {\widehat d}\to {\widehat d}_ {\rm \,C} \bigg]
- {\tilde b}\sigma 
\int_{0}^{\infty}\! \! dz  \; z
\frac{{\widehat d_{\rm \,N}}(z) - {\widehat d}_ {\rm \,C}(z)}{\bar{g}^2(z)},
\label{ddiv}
\ee
The first term on the RHS  of Eq.(\ref{ddiv}) 
is simply the RHS of  Eq.(\ref{dcorn}),  with 
${\widehat d}\to {\widehat d}_ {\rm \,C}$; it clearly converges,
since ${\widehat d}_ {\rm \,C}$ is (by construction) a canonical solution.
The second term receives an
appreciable contribution only in the  low
momenta region, i.e. from $0$  to $0.2 \;\mbox{GeV}^{\,2}$,
vanishing very rapidly in the UV, due to the
perfect  agreement  found between  both  curves  for  the range  $[0.2
\;\mbox{GeV}^{\,2},\Lambda_{\rm{UV}}]$.

\begin{figure}[ht]
\hspace{-1.5cm}
\includegraphics[scale=1.0]{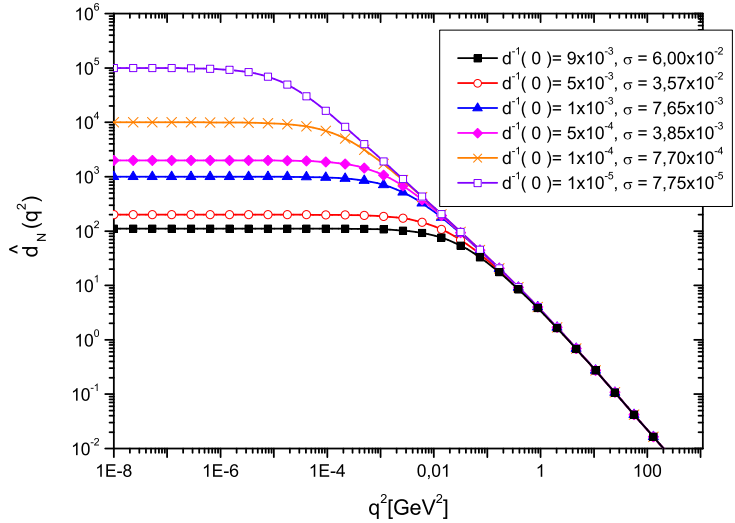}
\caption{Results for  ${\widehat d}(q^2)$ fixing  different values for
${\widehat  d}^{\,-1}(0)$ (all in $\mbox{GeV}^{\,2}$) in  a log-log
plot.  All   solutions  presented  satisfy  the   condition  given  by
Eq.(\ref{ddiv}). Their respective values for $\sigma$ are described on
its legend.}
\label{d4}
\end{figure}

 Other non-canonical solutions with different  ${\widehat d}^{\,-1}(0)$ are
plotted on Fig.~{\ref{d4}},
and  their  respective  values   of  $\sigma$  are  reported  in  the
legend. Notice  that, as the value of  ${\widehat d}(0)$ increases, 
the IR plateau becomes narrower.
       
%%%%%%%%%%%%%%%%%%%%%%%%Fig.13 %%%%%%%%%%%%%%%%%%%%%%%%%%%%%%%%%%%%%
%                    plot  - all non canonical propagators 
%%%%%%%%%%%%%%%%%%%%%%%%%%%%%%%%%%%%%%%%%%%%%%%%%%%%%%%%%%%%%%%%%%
%%%%%%%%%%%%%%%%%%%%%%%%%%%%%%%%%%%%%%%%%%%%%%%%%%%%%%%%%%%%%%%%%%% 

\begin{figure}[ht]
\hspace{-1.5cm}
\includegraphics[scale=1.0]{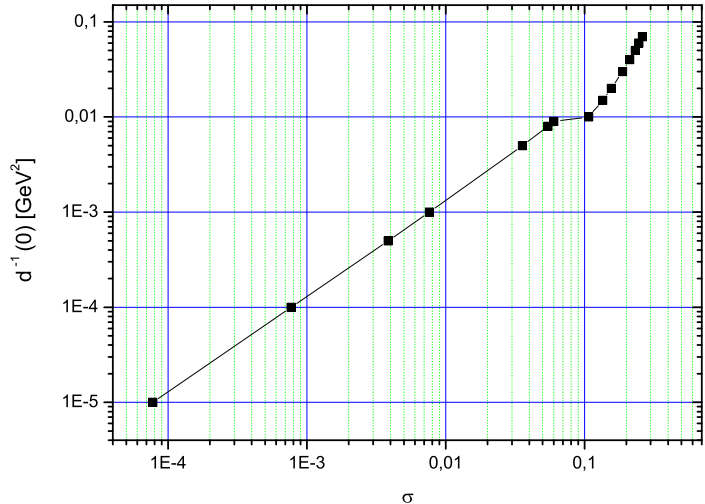}
\caption{ ${\widehat d}^{\,-1}(0)$ as function of $\sigma$.}
\label{d5}
\end{figure}

Finally, the  dependence of  ${\widehat d}^{\,-1}(0)$ on  $\sigma$ for
all cases presented here is  shown on Fig.~\ref{d5}, in a log-log
plot. The results clearly show a linear behavior for smaller values of
${\widehat   d}^{\,-1}(0)$,   whereas   for   values   of   ${\widehat
d}^{\,-1}(0)\geq 0.01 \;\mbox{GeV}^{\,2}$ the growth is exponential.

%%%%%%%%%%%%%%%%%%%%%%%%Fig.14 %%%%%%%%%%%%%%%%%%%%%%%%%%%%%%%%%%%%
%                plot - d0 versus sigma
%%%%%%%%%%%%%%%%%%%%%%%%%%%%%%%%%%%%%%%%%%%%%%%%%%%%%%%%%%%%%%%%%%
%%%%%%%%%%%%%%%%%%%%%%%%%%%%%%%%%%%%%%%%%%%%%%%%%%%%%%%%%%%%%%%%%%% 

\newpage 
\section{Discussion and Conclusions}
\label{Sect:Concl}

In this article we have taken a closer look at various issues relevant
to  the study  of gluon  mass  generation through  SD equations.   The
emphasis of our analysis has focused on the following points: (i) The
gauge-invariant truncation  scheme based on the  PT-BFM formalism, and
the  possibility it  offers  in implementing a self-consistent  first
approximation, omitting  ghost contributions without  compromising the
transversality  of  the  gluon  self-energy.  (ii)  The  necessity  of
introducing massless  poles in the form  of the vertex used  in the SD
equation,   and  the   role  of   the  seagull   graph   in  enforcing
transversality   (iii)   A  method   for   regulating  the   resulting
seagull-type  integrals based on  dimensional regularization,  and the
asymptotic properties  that the  solutions must display  in the  UV in
order  for  this regularization  to  work.   (iv)  We have  derived  a
linearized  version  of  the  SD  equation that  originates  from  the
one-loop dressed approximation, in the absence of ghost loops.  (v) We
have introduced  a phenomenological  form for the  three-gluon vertex,
containing  massless poles  associated with two  different tensorial
structures.  (vi)  The resulting equation has  been solved numerically
and two types of qualitatively distinct solutions have been found.

It should be obvious  from the present work that  the role  of  the three-gluon
vertex is  absolutely central; in particular, the  question of whether
or  not it contains massless  poles  is a determining factor  when
looking for finite solutions.  As  we have mentioned already, the form
of  the vertex  employed  here attempts  to  capture some  of the  main
features, such  as the impact of  the poles, and  the possible interplay
between  poles originating  from different  Lorentz structures (see
discussion after (\ref{D0})). Needless to  say, an in-depth study of the form
of  the   vertex  is  indispensable  for   further  substantiating the
appearance  of  IR-finite  solutions.  At  the  moment  we  only  have
indications  from the  study of  SD equation  and  lattice simulations
(albeit for  the three-gluon vertex  in the conventional  Landau gauge)
that the IR behavior is indeed singular.
In addition, one must explore the possibility of improving the 
gauge-technique inspired Ans\"atze employed 
for the vertex, in the spirit of~\cite{King:1982mk},
in an attempt to correctly 
incorporate the required asymptotic behavior into the 
SD equation (see Sec.~\ref{Sect:GenCon}). Given the rich tensorial structure of the 
three-gluon vertex~\cite{Binger:2006sj}, 
such a task is expected to be technically rather demanding.

It would be interesting to further scrutinize the viability of the new
class  of solutions  encountered  in Sec.~\ref{Sect:NumAn}. Such solutions  are
particularly interesting,  because they could in  principle overcome a
well-known  difficulty related  to  the breaking  of chiral  symmetry.
Specifically,  the  class   of  solutions  displaying  freezing,  when
inserted into standard  forms of the gap-equation for  the quarks, are
not able  to trigger  non-trivial solution, because  they do  no reach
high   enough   values   in   the   IR  to   overcome   the   critical
coupling~\cite{Papavassiliou:1991hx}.    Instead,  the  new   type  of
solutions rises sharply  in the deep IR, reaching  values that could in
principle break chiral symmetry, while for intermediate IR momenta it
coincides with a canonical massive  type of solution, that seems to be
favored by  phenomenology~\cite{Aguilar:2002tc} .
Of course, it could well happen that  
these  solutions are particular  to the  linearized
equation, and do not  survive a non-linear analysis.

In recent years  the picture that has emerged through  the study of SD
equations  in the conventional  Landau gauge  is characterized  by the
so-called ``ghost  dominance''~\cite{vonSmekal:1997is, Atkinson:1997tu}.  
In particular,  the gluon self-energy
has    the   form    $\Delta_{\,\mu\nu}(q^2)   =    \Delta(q^2)   {\rm
P}_{\mu\nu}(q)$,  with   $\Delta(q^2)=  Z(q^2)/q^2$,  and   the  ghost
propagator  is  $D(q^2)= -  C(q^2)/q^2$.   Assuming  that the  dressed
ghost-gluon vertex is finite in  the IR, the SD equations yields
for $q^2 \rightarrow  0$, $C(q^2) =A \, q^{-2\kappa}$,  $Z(q^2) = B \,
q^{\,4\kappa}$, where  the constants $A$  and $B$ depend  on $\kappa$.
The value  of $\kappa$ depends on  the details of the  dressing of the
gluon-ghost vertex  at small momenta;  SD and lattice studies  seem to
restrict  it within  the  range $0.5\le\kappa<0.6$.   For the  special
value  $\kappa=0.5$  one obtains  a  finite gluon  propagator,
$\Delta(0)=B$, whereas the ghost  propagator diverges as $D(q^2) \to -
A (q^2)^{-3/2}$.  In addition,  by virtue of the identity $\tilde{Z_1}
=  Z_g  Z_3^{1/2}\tilde{Z_3}=1$,  valid  in the  Landau  gauge,  where
$\tilde{Z_1}$,  $Z_g$,  $Z_3$ and  $\tilde{Z_3}$  are the  gluon-ghost
vertex,   coupling,   gluon   and  ghost   renormalization   constants
respectively, one  concludes that the product  $g^2Z(q^2) C^2(q^2)$ is
RG-invariant,   and  can   be   adopted  as   a   definition  of   the
non-perturbative   running  coupling,   i.e.   $\alpha(q^2)=(g^2/4\pi)
Z(q^2)C^2(q^2)$.  Then  it is clear that the  $\alpha(q^2)$ so defined
has  an IR  fixed point, $\alpha(0)=  (g^2/4\pi) A^2\,B  \,$, 
regardless of  the value  of $\kappa$.
The  actual value  of $\alpha(0)$
depends on $\kappa$, through the implicit dependence of $A$ and $B$ on
it; for  values of $0.5\le\kappa<0.6$, one  obtains $2.5 \le\alpha(0)<
3$.     Evidently,   the    (dimensionfull)    RG-invariant   quantity
$\widehat{d}(q^2)$ studied in this article displays a qualitatively similar
behavior to that found for the (dimensionless) running coupling in the
``ghost-dominance''  picture, namely IR  finiteness.  Therefore,
despite the  difference in the intermediate steps  and the terminology
employed,  the  physics  captured  by  both  pictures  appears  to  be
compatible.  Notice also  that, from the practical point  of view, the
method presented  here has the advantage  of setting up  a SD equation
directly  for the RG-invariant  object of  interest; thus,  the result
obtained  does  not  depend  on  the exactness  with  which  a  subtle
cancellation in the  ratio of two quantities, one  tending to zero and
one to infinity,  is realized.  This last comment  may be particularly
relevant in the context of  lattice simulations, where the above ratio
is  studied numerically;  clearly, a  small deviation  from  the exact
cancellation (due  to residual  volume dependences, for  instance) may
lead to  serious qualitative discrepancies.   It is also  important to
mention  that,  whereas the  confinement  mechanism  within the
``ghost-dominance'' description is attributed
to  the divergence  of  the ghost propagator~\cite{Gribov:1977wm}, 
in the picture
where the gluon is  massive the origin of confinement is
the condensation 
of vortices~\cite{Cornwall:1979hz,Zakharov:2005cg}.
Actually, recent investigations advocate interesting
connections between
the center-vortex picture and the Gribov-horizon scenario
\cite{Gattnar:2004bf}.

As mentioned in the Introduction,  in the PT-BFM scheme the omission of
the ghost loop does not interfere with gauge-invariance, but it might
alter the actual  form of the gluon self-energy.  Therefore, the study
of  the  gluon-ghost  system   should  be  eventually  considered.  Of
particular relevance in such a  study is the nature of the gluon-ghost
vertices involved;  in fact,  in the PT-BFM  scheme there will  be two
such  vertices:  ${\g}_{\nu}$,   appearing  in  Fig.~\ref{group_b}, and
${\gb}_{\beta}$, which  will appear in  the SD equation for  the ghost
propagator. The  IR behavior of ${\gb}_{\beta}$ (in  the Landau gauge)
is currently under  investigation on the lattice \cite{Bloch:2003sk}; 
it  would clearly be
important to settle the issue of whether it is divergent or finite.

Last  but  not least,  the  theoretical  situation  concerning the  SD
equations within  the PT-BFM  scheme merits further  intense scrutiny.
First  of all,  the correspondence  between the  PT and  BFM  has been
established  perturbatively  to all  orders,  but  no analogous  proof
exists  non-perturbatively, i.e.   at the  level of  the  SD equations
themselves.  In  this article  we have {\it  assumed} that  the PT-BFM
correspondence persists non-perturbatively.   A preliminary study in a
simplified    context   (scalar    QED)   fully    corroborates   this
assumption~\cite{sqed}  ; however,  the actual  realization  is highly
non-trivial, and  its generalization to  QCD deserves a thorough
analysis.  In the same context, the consequences of the second crucial
ingredient,   namely  the  substitution   of  quantum   quantities  by
background ones inside the  loops, are virtually unexplored.  One must
study in detail, preferably in  the context of the toy model mentioned
above, the viability and  self-consistency of this procedure.  We hope
to be able to undertake some of these tasks in the near future.

\begin{acknowledgments}
This research was supported by Spanish MEC under the grant FPA 2005-01678 and by  Funda\c{c}\~{a}o de Amparo  \`{a} Pesquisa do  Estado de
S\~{a}o  Paulo (FAPESP-Brazil)  through  the grant 05/04066-0.
\end{acknowledgments}

\appendix*
\section{Feynman rules in the BFM}

In this Appendix we list for completeness the Feynman rules appearing in~\cite{Abbott:1980hw}.
%
%%%%%%%%%%%%%%%%%%%%%%%%%%%%%%%%%%%%%%%%%%%%%%%%%%%%%%%
%%%%%%%%%%%Numbering as A-1, A-2
\setcounter{equation}{0}
\renewcommand{\theequation}{A-\arabic{equation}}
%%%%%%%%%%%%%%%%%%%%%%%%%%%%%%%%%%%%%%%%%%%%%%%%%%%%%%%

\noindent

%%%%%%%%%%%%%%%%%%%%%%%%%%%%%%%%%%%%%%%%%%%%%%%%%%%%%%%%%%
%           Fig. A1      gluon propagator                 %
%%%%%%%%%%%%%%%%%%%%%%%%%%%%%%%%%%%%%%%%%%%%%%%%%%%%%%%%%%
\vspace{1cm}
\begin{tabular}{ll}                            %%%%%BEGIN TABULAR
\begin{minipage}{2cm}
{\includegraphics[scale=0.7]{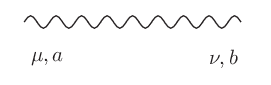}}
\end{minipage} &
\begin{minipage}{14.2cm}
\begin{equation}
\hspace{-2cm}
-i\left[g_{\mu\nu}-(1-\xi_Q)\frac{k_{\mu}k_{\nu}}{k^2}\right]\frac{\delta^{ab}}{k^2+i\epsilon}
\label{gl}
\end{equation}   
\end{minipage} \\
& \\
\end{tabular}
%%%%%%%%%%%%%%%%%%%%%%%%%%%%%%%%%%%%%%%%%%%%%%%%%%%%%%%%

%\noindent
%%%%%%%%%%%%%%%%%%%%%%%%%%%%%%%%%%%%%%%%%%%%%%%%%%%%%%%%%%
%        Fig.A2     ghost propagator                    %
%%%%%%%%%%%%%%%%%%%%%%%%%%%%%%%%%%%%%%%%%%%%%%%%%%%%%%%%%%
\begin{tabular}{ll}
\begin{minipage}{2cm}
{\includegraphics[scale=0.7]{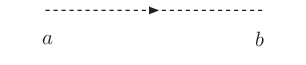}}
\end{minipage} &
\begin{minipage}{14.2cm}
%\begin{flushleft}
\begin{equation}
\frac{i\delta^{ab}}{k^2+i\epsilon}
\label{gh}
\end{equation}   
%\end{flushleft}
\end{minipage} \\
\end{tabular}
%%%%%%%%%%%%%%%%%%%%%%%%%%%%%%%%%%%%%%%%%%%%%%%%%%%%%%%%
\vspace{1cm}
%\noindent
%%%%%%%%%%%%%%%%%%%%%%%%%%%%%%%%%%%%%%%%%%%%%%%%%%%%%%%%%%
%        Fig.A3  Conventional 3-gluon vertex             %
%%%%%%%%%%%%%%%%%%%%%%%%%%%%%%%%%%%%%%%%%%%%%%%%%%%%%%%%%%
\begin{tabular}{ll}
\begin{minipage}{2cm}
{\includegraphics[scale=0.7]{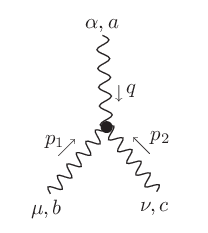}}
\end{minipage} &
\begin{minipage}{14.7cm}     %%%%size different
\begin{flushleft}
\begin{equation}
g f^{abc}\left[\left(p_1-q\right)_{\nu}g_{\mu\alpha}+\left(q -p_2\right)_{\mu}g_{\nu\alpha} + \left(p_2-p_1\right)_{\alpha}g_{\mu\nu}\right]
\label{3g_vertex}
\end{equation} 
\end{flushleft}
\end{minipage}\\
\end{tabular}                                             
%%%%%%%%%%%%%%%%%%%%%%%%%%%%%%%%%%%%%%%%%%%%%%%%%%%%%%%%
\vspace{1cm}
\noindent
%%%%%%%%%%%%%%%%%%%%%%%%%%%%%%%%%%%%%%%%%%%%%%%%%%%%%%%%%%
%        Fig. A4  Background 3-gluon vertex              %
%%%%%%%%%%%%%%%%%%%%%%%%%%%%%%%%%%%%%%%%%%%%%%%%%%%%%%%%%%
\begin{tabular}{ll}
\begin{minipage}{2cm}
{\includegraphics[scale=0.7]{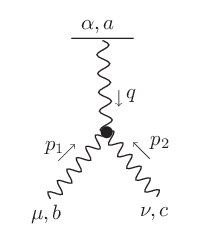}}
\end{minipage} &
\begin{minipage}{14.2cm}
\begin{flushright}
\begin{equation}
g f^{abc}\left[\left(p_1-q+\frac{1}{\xi_Q}p_2\right)_{\nu}g_{\mu\alpha}+\left(q -p_2-\frac{1}{\xi_Q}p_1\right)_{\mu}g_{\nu\alpha} + \left(p_2-p_1\right)_{\alpha}g_{\mu\nu}\right] \!\!
\label{bfm_3g}
\end{equation}
\end{flushright}
\end{minipage} \\
\end{tabular}
%%%%%%%%%%%%%%%%%%%%%%%%%%%%%%%%%%%%%%%%%%%%%%%%%%%%%%%%
\vspace{1cm}
\noindent
%%%%%%%%%%%%%%%%%%%%%%%%%%%%%%%%%%%%%%%%%%%%%%%%%%%%%%%%%%
%        Fig.A5  Conventional 4-gluon vertex and           %
% Background 4-gluon vertex (1 leg Background 3 legs Quantum)%
%%%%%%%%%%%%%%%%%%%%%%%%%%%%%%%%%%%%%%%%%%%%%%%%%%%%%%%%%%
\begin{tabular}{ll}
\begin{minipage}{2cm}
{\includegraphics[scale=0.7]{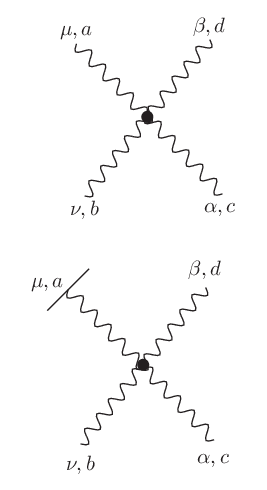}}
\end{minipage} &
\begin{minipage}{14.2cm}
\begin{flushright}
\begin{eqnarray}
&&-ig^2\bigg[
f^{abx} f^{xcd}\left(g_{\mu\alpha}g_{\nu\beta}-g_{\mu\beta}g_{\nu\alpha}
\right) \bigg. \nonumber \\
&& \bigg. +  f^{adx} f^{xbc}\left( g_{\mu\nu}g_{\alpha\beta}-g_{\mu\alpha}g_{\nu\beta}
 \right) \bigg.  \nonumber \\
&& + \bigg. f^{acx} f^{xbd}
\left(g_{\mu\nu}g_{\alpha\beta}-g_{\mu\beta}g_{\nu\alpha}\right) \bigg]
\label{4gluon}
\end{eqnarray}
\end{flushright}
\end{minipage}  \\
\end{tabular}
%%%%%%%%%%%%%%%%%%%%%%%%%%%%%%%%%%%%%%%%%%%%%%%%%%%%%%%%
\vspace{1cm}
\noindent
%%%%%%%%%%%%%%%%%%%%%%%%%%%%%%%%%%%%%%%%%%%%%%%%%%%%%%%%%%
%        Fig.A6  Background 4-gluon vertex                %
%%%%%%%%%%%%%%%%%%%%%%%%%%%%%%%%%%%%%%%%%%%%%%%%%%%%%%%%%%
\begin{tabular}{ll}
\begin{minipage}{2cm}
{\includegraphics[scale=0.7]{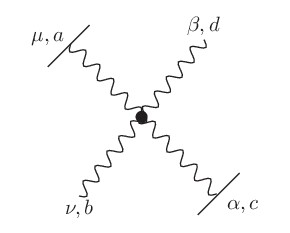}}
\end{minipage} &
\begin{minipage}{14.2cm}
\begin{flushright}
\begin{eqnarray}
&&-ig^2\bigg[
f^{abx} f^{xcd}\left(g_{\mu\alpha}g_{\nu\beta}-g_{\mu\beta}g_{\nu\alpha}
+ \frac{1}{\xi_Q}g_{\mu\nu}g_{\alpha\beta}\right) \bigg. \nonumber \\
&& \bigg. +  f^{adx} f^{xbc}\left( g_{\mu\nu}g_{\alpha\beta}-g_{\mu\alpha}g_{\nu\beta}
- \frac{1}{\xi_Q}g_{\mu\beta}g_{\nu\alpha} \right) \bigg.  \nonumber \\
 &&+ \bigg. f^{acx} f^{xbd}
\left(g_{\mu\nu}g_{\alpha\beta}-g_{\mu\beta}g_{\nu\alpha}\right) \bigg]
\label{bfm4gluon}
\end{eqnarray}
\end{flushright}
\end{minipage} \\
\end{tabular}
%%%%%%%%%%%%%%%%%%%%%%%%%%%%%%%%%%%%%%%%%%%%%%%%%%%%%%%%
\vspace{1cm}
\noindent
%%%%%%%%%%%%%%%%%%%%%%%%%%%%%%%%%%%%%%%%%%%%%%%%%%%%%%%%%%
%        Fig.A7  Background gluon-ghost vertex             %
%%%%%%%%%%%%%%%%%%%%%%%%%%%%%%%%%%%%%%%%%%%%%%%%%%%%%%%%%%
\begin{tabular}{ll}
\begin{minipage}{2cm}
{\includegraphics[scale=0.7]{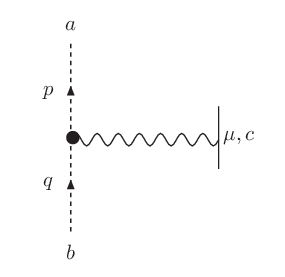}}
\end{minipage} &
\begin{minipage}{14.2cm}
\begin{flushright}
\begin{equation}
-g f^{abc}(p+q)_{\mu}
\label{bfm_ghostv1}
\end{equation}
\end{flushright}
\end{minipage}  \\
\end{tabular}
%%%%%%%%%%%%%%%%%%%%%%%%%%%%%%%%%%%%%%%%%%%%%%%%%%%%%%%%
\vspace{1cm}
\noindent
%%%%%%%%%%%%%%%%%%%%%%%%%%%%%%%%%%%%%%%%%%%%%%%%%%%%%%%%%%
%        Fig.A8  Conventional gluon-ghost vertex           %
%%%%%%%%%%%%%%%%%%%%%%%%%%%%%%%%%%%%%%%%%%%%%%%%%%%%%%%%%%
\begin{tabular}{ll}
\begin{minipage}{2cm}
{\includegraphics[scale=0.7]{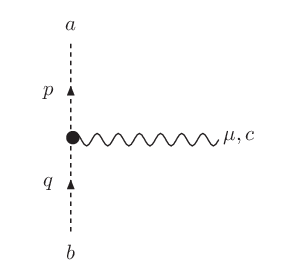}}
\end{minipage} &
\begin{minipage}{14.2cm}
\begin{flushright}
\begin{equation}
-g f^{abc}p_{\mu}
\label{bfm_ghostv2}
\end{equation}
\end{flushright}
\end{minipage}  \\
\end{tabular}
%%%%%%%%%%%%%%%%%%%%%%%%%%%%%%%%%%%%%%%%%%%%%%%%%%%%%%%%
\vspace{1cm}
%%%%%%%%%%%%%%%%%%%%%%%%%%%%%%%%%%%%%%%%%%%%%%%%%%%%%%%%%%
%        Fig.A9  Background  4-fields vertex               %
% (1 background gluon, 1 quantum gluon and 2 quantum ghost)      
%%%%%%%%%%%%%%%%%%%%%%%%%%%%%%%%%%%%%%%%%%%%%%%%%%%%%%%%%%
\begin{tabular}{ll}
\begin{minipage}{2cm}
{\includegraphics[scale=0.7]{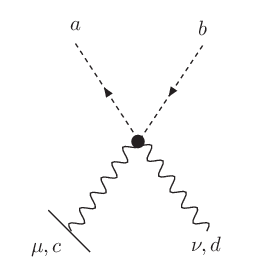}}
\end{minipage} &
\begin{minipage}{14.2cm}
\begin{flushright}
\begin{equation}
-ig^2 f^{acx}f^{xdb}g_{\mu\nu}  
\label{bfm_v1}
\end{equation}
\end{flushright}
\end{minipage}  \\
%\end{tabular}
%%%%%%%%%%%%%%%%%%%%%%%%%%%%%%%%%%%%%%%%%%%%%%%%%%%%%%%%
\vspace{1cm}
%%%%%%%%%%%%%%%%%%%%%%%%%%%%%%%%%%%%%%%%%%%%%%%%%%%%%%%%%%
%        Fig10  Background  4-fields vertex              %
%      (2 background gluon and 2 quantum ghost)          %
%%%%%%%%%%%%%%%%%%%%%%%%%%%%%%%%%%%%%%%%%%%%%%%%%%%%%%%%%%
%\begin{tabular}{ll}
\begin{minipage}{1cm}
{\includegraphics[scale=0.7]{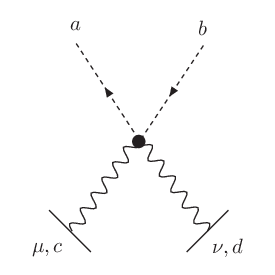}}
\end{minipage} &
\begin{minipage}{14.2cm}
\begin{flushright}
\begin{equation}
-ig^2 g_{\mu\nu}\left(f^{acx}f^{xdb}+f^{adx}f^{xcb}\right)
\label{bfm_v2}
\end{equation}
\end{flushright}
\end{minipage}
\end{tabular}
%%%%%%%%%%%%%%%%%%%%%%%%%%%%%%%%%%%%%%%%%%%%%%%%%%%%%%%%
\vspace{1cm}

%\end{appendix}

%\newpage


\begin{thebibliography}{99}
\bibitem{foot1}
It should be clear that the seagull contribution cannot be 
``projected out'' and discarded, since it  
forms part of the scalar cofactor multiplying 
the projection operator ${\rm P}_{\mu\nu}(q)$.
The only reason why this can be done some times in the literature 
is because the transversality of the answer has been 
compromised at an earlier stage, due to some incomplete approximation.
\end{thebibliography}
\end{document}